\documentclass[smallextended]{svjour3}       
\smartqed  

\usepackage{graphicx,psfrag,color}
\usepackage{dcolumn}
\usepackage{bm}
\usepackage{mathbbol}

\def\tr{{\rm tr}\; }

\begin{document}


\title{Teleportation-Based Continuous Variable Quantum Cryptography}

\author{F. S. Luiz  \and Gustavo Rigolin}


\institute{F. S. Luiz \at
           Departamento de F\'{i}sica,
	   Universidade Federal de S\~ao Carlos, S\~ao Carlos, SP 13565-905,
	   Brazil          
          \and
           Gustavo Rigolin \at
           Departamento de F\'{i}sica,
	   Universidade Federal de S\~ao Carlos, S\~ao Carlos, SP 13565-905,
	   Brazil \\  
           \email{rigolin@ufscar.br}  
}

\date{Received: date / Accepted: date}

\maketitle

\begin{abstract}
We present a continuous variable (CV) quantum key distribution (QKD) scheme based on the 
CV quantum teleportation of coherent states
that yields a raw secret key made up of discrete
variables for both Alice and Bob. 
This protocol preserves the efficient detection schemes  
of current CV technology (no single-photon detection techniques) and, at the same time,  
has efficient error correction and privacy amplification schemes due to the binary modulation of the key.  
We show that for a certain type of incoherent attack it is  
secure for almost any value of the transmittance of the optical line used by 
Alice to share entangled two-mode squeezed states with Bob 
(no 3 dB or $50\%$ loss limitation characteristic of beam splitting attacks).
The present CVQKD protocol works deterministically (no postselection needed) with efficient direct reconciliation techniques 
(no reverse reconciliation) in order to generate a secure key and beyond the $50\%$ loss case at the
incoherent attack level.
\end{abstract}






\section{Introduction}

Currently, the only absolutely secure way
through which two parties (Alice and Bob) can, at least theoretically, secretly share a random sequence of bits (key) 
is given by quantum cryptography, whose security is guaranteed by the validity of the laws
of quantum mechanics \cite{Ben84}. This secret key is the most important ingredient in the implementation of classical 
cryptography protocols, such as the one-time pad, which are provably secure if the key is only known by Alice and Bob.

The original QKD protocols are based on single photons (``discrete'' states), requiring 
photon-counting techniques to their implementation \cite{Ben84,Gis02}.  
However, single photon detectors are not as efficient and fast (short response time) as standard 
telecommunication PIN photodiodes used to detect bright light (many photons) \cite{Gis02}. 
In quantum mechanics these bright quantum states are described
by the quadratures of a mode of the quantized electromagnetic field and are also known as CV states due to the continuum spectrum of the
quadratures.  
In order to explore the efficient and fast measurement schemes for such states (homodyne or heterodyne detection), QKD protocols
based on several types of CV states and strategies were proposed \cite{Ral99,Cer01,Gro02,Leu02,Gro03,Nam03,Wee04,Hei06,Pir08,Pat09,Lev09,Leu10,Mad12,Pir14,Bor16}. 
They are all called CVQKD protocols \cite{Bra05} and are considered theoretically secure \cite{Gro04}.

The quantum resources of the first CVQKD protocols \cite{Ral99,Cer01}, 
whose security was equivalent to discrete QKD protocols, 
were either single-mode squeezed states, sent from Alice to Bob, or 
two-mode entangled squeezed states shared between them. In these early schemes 
the secret key was encoded either in binary alphabets composed of two different
states (discrete modulation) \cite{Ral99} or in states with real and imaginary quadratures \cite{Cer01}
chosen from Gaussian distributions (continuous modulation)\footnote{A discrete variable QKD scheme is based on the use of 
qubits or qudits (finite dimensional Hilbert spaces) while a CVQKD scheme employs physical systems described by
infinite dimensional Hilbert spaces (such as coherent and squeezed states). Note, however, that in CVQKD protocols
the key can be modulated using either discrete or continuous alphabets/variables \cite{Bra05}. In the present protocol
we use a discrete alphabet of coherent states to modulate the key and a two-mode squeezed state to teleport the coherent state from
Alice to Bob. This is why we call our protocol a teleportation-based CVQKD scheme.}. 
An important development of CVQKD appeared in \cite{Gro02}, where it was shown
that coherent states are equally secure to generate a secret key between Alice and Bob if one uses a Gaussian continuous modulation and if the transmission
loss from Alice to Bob does not exceed $50\%$. Subsequently, in \cite{Leu02} it was shown that if Bob accepts only certain measurement outcomes (postselection)
to generate the key, or if Alice and Bob employ reverse reconciliation techniques \cite{Gro03}, they can surpass the $50\%$ loss threshold. 
Also, by employing at the
same time reverse reconciliation and postselection one gets the greatest secure key rates \cite{Hei06}.

A reconciliation technique is an error correction scheme implemented at the end of the protocol by Alice and Bob, 
in which they execute a set of tasks in order to agree on a common sequence of bits. This process is called direct if Alice, who sends the quantum states,
communicates classically with Bob, who then processes his data using a predetermined algorithm to agree with Alice's random sequence of bits. 
Reverse reconciliation is the opposite scenario, where Bob communicates with Alice, who now manipulates
her data in order to share a common key with Bob. So far, there is no CVQKD protocol that is secure for 
any value of loss that uses only direct reconciliation and no postselection.

In this article we show a different way to do CVQKD that is secure against individual attacks
for losses close to 100$\%$ without resorting to either 
reverse reconciliation or postselection, the standard solutions to make a CVQKD protocol work securely for
losses greater than $50\%$. 
Since this protocol works deterministically (no postselection) and uses a discrete
modulation for the key, it achieves fairly high key rates over long distances, even assuming the usual 
conservative reconciliation efficiencies for CV protocols based on binary modulated keys \cite{Lev09}.
Apart from its possible practical significance, this protocol also adds to our fundamental understanding of CVQKD since it is based on
the active use of CV teleportation protocols \cite{Bra98}, opening up alternative ways to understand the security of CVQKD as well as different
routes for future unconditional security proofs.

\begin{figure}[!ht]
\begin{center}
\includegraphics[width=12cm,height=6.37cm]{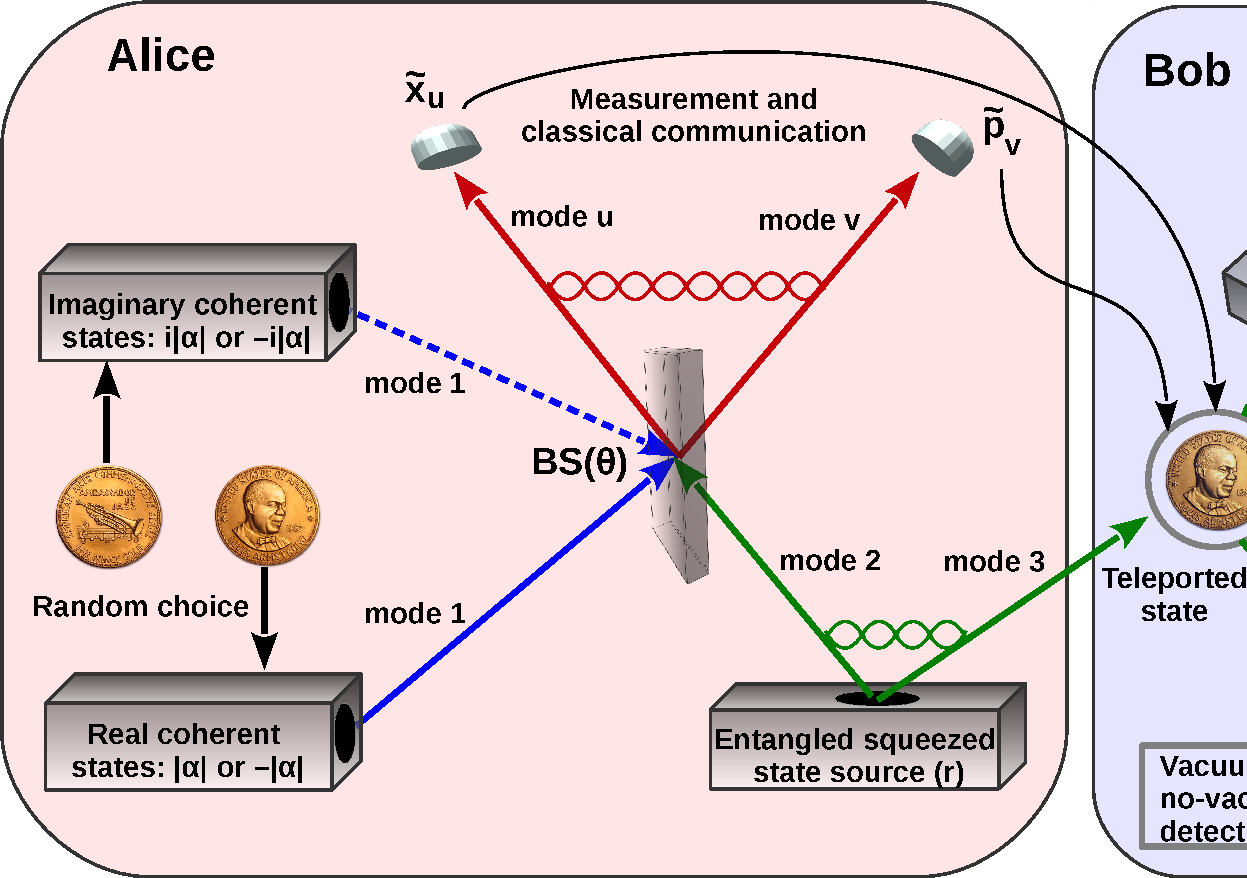}
\end{center}
\caption{\label{fig_esquema}  Schematic representation of the teleportation-based CVQKD protocol.
The encoding of the binary key on which Alice and Bob agree is 
$\{|-\alpha\rangle,|\alpha\rangle,|-i\alpha\rangle,|i\alpha\rangle\}=\{0,1,0,1\}$, where $\alpha$ is a real
number. See text and Appendix \ref{C} for details. 
}
\end{figure}

Following \cite{Gor10,Lui14}, the main idea behind the present teleporta\-tion-based CVQKD scheme is the active use of the finite resources (finite squeezing) 
inherently associated to the CV teleportation protocol, combined with
the knowledge of the pool of coherent states with Alice to be teleported to Bob \cite{Lui14}. It is by properly making use of these two
pieces of information that we can build a protocol furnishing high key rates even in a scenario with high losses, 
turning the finiteness of squeezing into 
an advantage. 
Indeed, the CV teleportation protocol is not simply employed as an alternative to the direct sending 
of the states with Alice to Bob, as required by the aforementioned standard CVQKD protocols, where the greater the entanglement of
the channel the more a flawless teleportation is achieved with subsequent higher key 
rates\footnote{Note that the goals of the standard CV teleportation protocol \cite{Bra98} as well as the one of Ref. \cite{Lui14} 
are not a secure transmission of quantum states. 
The generalized CV teleportation protocol of Ref. \cite{Lui14} is used here as a tool to the development of the 
present CVQKD scheme. Without the present modifications, the protocols given in Refs. \cite{Bra98,Lui14} are not able to achieve
a secure transmission of quantum states.}.
In the present protocol, however, less entanglement means more efficiency (see Appendix \ref{E}),
since we show that for a lossy transmission the amount of entanglement (squeezing) maximizing the key rate is finite, dependent
on the level of loss, and on the coherent states chosen for encoding the key. In other words, 
the maximally entangled (infinitely squeezed) 
channel connecting Alice and Bob is not the one yielding the greatest key rate.

\section{The protocol}

Let start describing the protocol (figure \ref{fig_esquema}), whose 
main ingredient is the modified CV teleportation protocol presented in \cite{Lui14}, where Bob can get an output state at the end of the 
teleportation nearly identical to the input state, even for low squeezing, if Alice and Bob know  
the set of input states to be teleported. (See Appendix \ref{B} for a self-contained presentation of the modified CV teleportation
protocol.) 
To achieve that Alice has to 
modify her beam splitter (BS) transmittance and Bob has to modify
the displacement $\hat{D}_k(\lambda)=e^{\lambda \hat{a}_k^\dagger - \lambda^*\hat{a}_k}$ on his mode, from those given by
the original CV teleportation protocol \cite{Bra98}, according to the pool of input states. 
Here $\hat{a}_k$ ($\hat{a}_k^\dagger$) is
the annihilation (creation) operator of mode $k$ with quadratures $\hat{x}_k=(\hat{a}_k+\hat{a}_k^\dagger)/2$
and $\hat{p}_k=(\hat{a}_k-\hat{a}_k^\dagger)/2i$ and commutation relation $[\hat{x}_k,\hat{p}_k]=i/2$. 
%
%

The present teleportation-based CVQKD protocol works as follows. 
Alice divides her pool of coherent states into two sets, $\{|\alpha\rangle,|-\alpha\rangle\}$ and 
$\{|i\alpha\rangle,|-i\alpha\rangle\}$, 
which we respectively call real and imaginary basis ($\alpha>0$). 
Alice and Bob agree beforehand on the following binary encoding \cite{Nam03} in order
to associate from each coherent state a bit value to the key:
$\{|-\alpha\rangle,|-i\alpha\rangle\} \rightarrow 0$ and $\{|\alpha\rangle,|i\alpha\rangle\} \rightarrow 1$. 
At each run of
the protocol, Alice randomly chooses between the real and imaginary basis and then randomly picks one of the two 
states belonging to the chosen basis. Let us generically call this state by $|\varphi\rangle$, which is teleported to Bob by means of 
a two-mode squeezed state $|\psi_r\rangle$, with squeezing parameter 
$r$ \cite{Bra98,Fur07}. $|\psi_r\rangle$ is prepared by Alice, who keeps one of its mode and send the other to Bob. 
In order to finish her part in the teleportation, Alice combines her share of the entangled resource with $|\varphi\rangle$
in a BS with transmittance $\cos^2\theta$. 
After measuring the position and momentum quadratures of the modes outgoing the BS, Alice informs 
Bob of her measurement results ($\tilde{x}_u$ and $\tilde{p}_v$). 

\begin{figure}[!ht]
\begin{center}
\includegraphics[width=8cm]{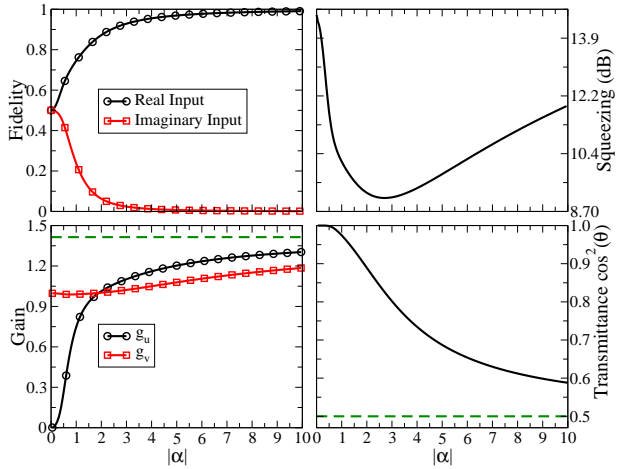}
\end{center}
\caption{\label{figFid} Optimized parameters giving the greatest (least) fidelity for a teleported real (imaginary) coherent state.
The optimal settings for the greatest (least) fidelity for an imaginary (real) input are obtained from the ones above by interchanging 
$g_u$ with $g_v$ and changing $\theta$ to $\pi/2 - \theta$. The squeezing remains unchanged. 
The dashed curves give the settings for the original CV teleportation protocol \cite{Bra98}.}
\end{figure}

Bob, who now knows the values of $\tilde{x}_u$ and $\tilde{p}_v$, randomly chooses between two possible types of displacements
$\hat{D}(\lambda)$ to implement on his mode ($\lambda = g_u\tilde{x}_u+ig_v\tilde{p}_v$), which we call
real and imaginary displacements. 
These different types of displacements are given by different pairs of gains $(g_u,g_v)$ and are optimized in the following sense.
The real (imaginary) displacement is such that Bob's state, $\hat{\rho}_B$, has the greatest fidelity possible with Alice's input if
she chose the real (imaginary) basis and the least fidelity if her choice was the imaginary (real) basis. 
Moreover, this is done such that the optimal $(g_u,g_v)$ do not depend on the sign of the teleported coherent state but only on its being a real
or imaginary state (see figure \ref{figFid} and the following paragraphs).  
Note that by fidelity we mean a quantity $F \in [0,1]$ that measures the similarity between 
two quantum states and in our case can be written as 
$F=\langle \varphi | \hat{\rho}_B|\varphi \rangle$, where $F=0$ for orthogonal states and $F=1$ for identical ones.

\begin{figure}[!ht]
\begin{center}
\includegraphics[width=8cm]{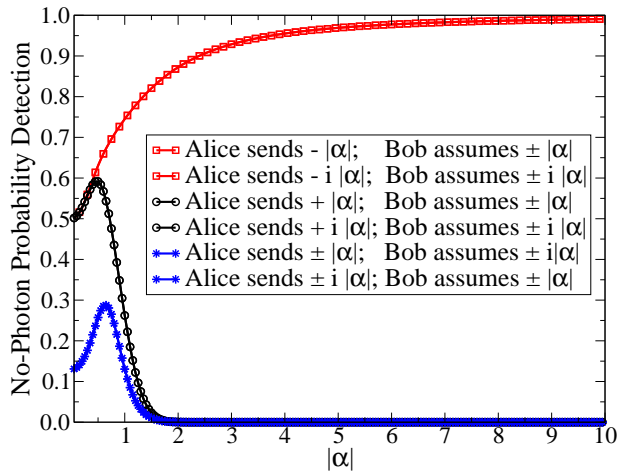}
\end{center}
\caption{\label{figProb} Probabilities for Bob detecting the vacuum state at the end of a run of the protocol 
if Alice and Bob use the optimal settings given in
figure \ref{figFid}. Whenever Bob (or Eve) assumes incorrectly the basis employed by Alice, he (she) cannot discern
between the two possible inputs (star/blue curves). Note that this fact resembles the key generation scheme of the BB84 protocol \cite{Ben84}
and is one of the reasons why this protocol is successful in establishing a 
secure key between Alice and Bob. 
}
\end{figure}

The next step of the protocol
consists in Bob once again displacing his state. He applies $\hat{D}(\alpha)$ to his mode if he previously implemented the real displacement
or $\hat{D}(i\alpha)$ otherwise.  
The goal of this last displacement is to transform either the states $|-\alpha\rangle$ or $|-i\alpha\rangle$ to vacuum states or to
move farther from the vacuum the states $|\alpha\rangle$ or $|i\alpha\rangle$. One of these real (imaginary) states nearly describes 
$\hat{\rho}_B$ if Alice chose the real (imaginary) basis and Bob the real (imaginary) displacement in a given run of the protocol. 
After the last displacement Bob measures the intensity of his mode and associates the bit $0$ if he sees no light (vacuum state) or
the bit $1$ if he sees any light (see figure \ref{figProb}). Note that the previous step can be modified to any strategy aimed to discriminate between two 
coherent states, such as the measurement of the quadratures of $\hat{\rho}_B$ using homodyne detection.

Alice and Bob repeat the previous steps until they have enough data to check for an eavesdropper and still get a secure key long enough
for their purposes. After Alice finishing all teleportations and after Bob making all measurements, they use an authenticated classical channel
to disclose the following information. Alice reveals to Bob the basis 
used at each run of the protocol but not the state. Bob reveals to Alice the instances where he used the optimal values of $g_u$ and $g_v$ matching
the basis chosen by Alice. They discard the data where no matches occurred and use a sample of the remaining data 
to check for the parameters of the quantum channel (loss and noise) they previously determined or assumed and to check for security. Then  
they implement error correction techniques on the non disclosed data (reconciliation stage) in order to agree on the random sequence of zeros and ones and,
subsequently, generate the final secret key via standard privacy amplification techniques (classical algorithms
devised to enhance the privacy of a shared random sequence of data). 

As mentioned above, a key feature of the present protocol is the fact that the optimal $(g_u,g_v)$ 
can be chosen such that they do not depend on the sign of the teleported coherent state, depending only
on the state being real or imaginary. 
We can see that this can be done by looking at the functional form of $F$ after a single run of the protocol. 
Assuming, for definiteness, we are dealing with real coherent states we have 

\begin{equation}
\hspace{-1.5cm}F = h_1(r,\theta)\exp[f_1(\tilde{p}_v,\tilde{x}_u,g_v,g_u,r,\theta)+2\alpha \tilde{x}_u f_2(g_u,r,\theta)+\alpha^2 f_3(r,\theta)], 
\label{extra1}
\end{equation}
where the functions $h_1$ and $f_j, j=1,2,3$, are given in Appendix \ref{C}, along with all the mathematical details 
needed to understand the present protocol. Looking at ($\ref{extra1}$) we see that $F$ depends on $\alpha$
only linearly and quadratically. This means that we can cancel the dependence of $F$ on the sign of $\alpha$ by eliminating 
the linear dependence on it. This is achieved by demanding that 
\begin{equation}
f_2(g_u,r,\theta)=0,
\label{extra2}
\end{equation}
which leads to 
\begin{equation}
g_u(r,\theta) = \frac{\sinh (2 r) \sin \theta}{\cosh^2r-\cos(2 \theta )\sinh^2r }.
\label{extra3}
\end{equation}
Furthermore, maximizing $F$ with respect to $g_v$ immediately gives
\begin{equation}
g_v(r,\theta) = \frac{2 \coth r \cos\theta}{\coth^2r+\cos(2 \theta )}.
\label{extra4}
\end{equation}
In figure \ref{figFid} we show the optimal values for these quantities,
where $r$ and $\theta$ are chosen such that we get the greatest fidelity
for a real teleported state and the least fidelity when teleporting an imaginary state.
It is worth mentioning that this protocol is very robust to fluctuations about those optimal 
values as we show in detail in Appendix \ref{C}. Also, the physical resources needed to implement 
the present protocol with reasonable key rates are already available, in particular the efficient production of 
two-mode squeezed states \cite{Fur07}, the main ingredient of the present protocol.

Before we proceed with the security analysis of this protocol, 
let us review what we have so far. 
We showed, first, that it is possible to choose optimal parameters maximizing the fidelity \textit{independently} of the sign of the teleported
coherent state and, second,  that this choice \textit{depends} on the coherent state being real or imaginary. 
Third, we also showed that Bob can discern which state Alice teleported if, and only if, he chooses the right displacement to implement on
his mode at the end of a single run of the protocol (see figure \ref{figProb}). Those three features reminds us of the 
working principles of the BB84 protocol \cite{Ben84}, where Bob can only obtain
the right bit in a given run of the protocol 
if he measures his qubit using the same basis employed by Alice to prepare it. In our case, the non-orthogonal basis 
of the BB84 protocol is related to the real and imaginary basis defined here; and the fact that the BB84 protocol only succeeds if Bob
chooses the right measurement basis is connected here to the fact that Bob must choose the right displacements $g_u$ and $g_v$ to succeed.

\section{Security analysis}

Let us move to the security analysis, where we deal with individual (incoherent) attacks only. 
The inter\-cept-resend attack, with an eavesdropper
Eve blocking Bob's share of the entangled state 
(mode 3 in figure \ref{fig_esquema}) and sending him a fake mode, 
is not as serious a threat as the BS attack we will be dealing with in what follows. 
This is true because Eve cannot know Alice's input with certainty \textit{before} sending Bob the fake mode. 
Indeed, Eve can only hope to know Alice's input by knowing which basis she used and this only happens 
\textit{after} Bob measures his mode.

The most serious incoherent attack to the present and all CVQKD schemes is the BS attack, in which Eve inserts a BS of transmittance $\eta$ 
in the optical line connecting Alice and Bob and operates on the signal reaching her ($1-\eta$) in the same way as Bob does with
his share of the signal ($\eta$). 
Note that the BS attack is equivalent to a lossy transmission where $1-\eta$ of the signal is lost to the environment. For direct
reconciliation \cite{Gro02,Leu02}, the secure key rate generated between Alice and Bob in the BS attack is 
\begin{equation}
K = \max\{0, \beta I_{AB}-I_{AE}\}=\{0,\Delta I\},
\end{equation}
where
$I_{AB}$ and $I_{AE}$ are the mutual information between Alice and Bob and Alice and Eve, respectively. $\beta$ is the reconciliation
efficiency and depends on the reconciliation software employed. For binary encodings that we use here it has a conservative value 
of $\beta \approx 80\%$ \cite{Lev09}. Since the present protocol and the BS attack are symmetric with respect the real and imaginary
states, in the following security analysis we consider only the case where Alice used the real basis and Bob and Eve the real displacement,
i.e., we assume Alice teleported either the coherent state $|\alpha\rangle$ or $|-\alpha\rangle$, with $\alpha$ real, and Bob and Eve correctly
guessed that Alice chose a real coherent state.

A direct calculation of the mutual informations gives (see Appendix \ref{D}),
\begin{eqnarray}
I_{AY} &=& 1 + \sum_{j=0}^1 [q_j^Y\log_2 q_j^Y + (1-q_j^Y)\log_2 (1-q_j^Y) \nonumber \\
&&-(1+q_j^Y-q_{\bar{j}}^Y)\log_2(1+q_j^Y-q_{\bar{j}}^Y)]/2,  
\end{eqnarray}
where $Y=B$ or $E$, $\bar{j}=0(1)$ if $j=1(0)$, and $q_j^Y$ is the unconditional (no postselection) probability of $Y$
to assign the bit $j$ to the key if Alice teleported the corresponding state that encodes the bit $j$.  
In the present case $q_0^Y$ means the probability of $Y$ to detect the vacuum state at the end of a run of the protocol 
if Alice teleports $|-\alpha\rangle$ while $q_1^Y$ is the probability of $Y$ to detect any light if she teleports
$|\alpha\rangle$. Note that $q_j^Y$ depends on $\eta$ and that $q_j^B(\eta)=q_j^E(1-\eta)$.
\begin{figure}[!ht]
\begin{center}
\includegraphics[width=8cm]{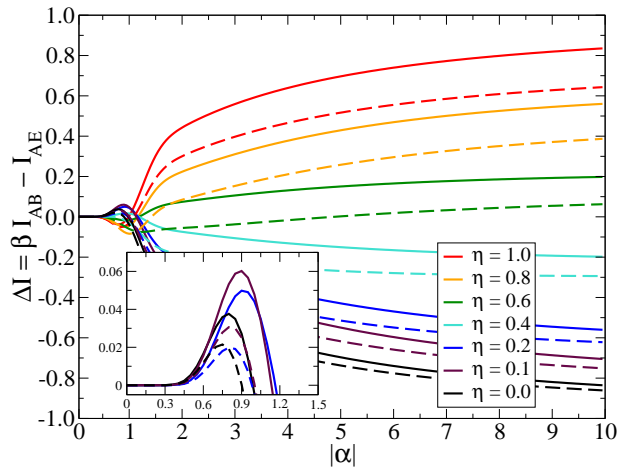}
\end{center}
\caption{\label{figKey} All plots: Solid lines mean $\beta =1.0$ and dashed lines $\beta=0.8$.
Main plot: For large values of $|\alpha|$ we have from top to bottom increasing (decreasing) loss ($\eta$).
Inset: $\eta =0.1$ ($90\%$ loss) for the greatest peaks of the solid and dashed curves while $\eta = 0.2$ ($80\%$ loss) 
for the corresponding lowest ones.
}
\end{figure}

In figure \ref{figKey} we plot $\Delta I$ for several values of loss employing the parameters shown in figure \ref{figFid}.
The inset shows that it is possible to choose a value of $\alpha$ such that for $\beta = 0.8$ and $90\%$ loss we get $K \approx 0.03$.
This value should be contrasted with those without excess noise in \cite{Leu02}, where by setting a perfect direct reconciliation
($\beta = 1$) and postselection one gets $K=0.007$ at $75\%$ loss, and with the ones in \cite{Hei06}, where above $80\%$ loss it is not 
possible to extract a secret key via direct reconciliation. In other words, we improve the key rate at about one order of magnitude 
even assuming more loss. To get such enormous gain in the key rate we need a squeezing of about $10$ dB.  

When the loss is exactly $100\%$, the protocol does not work since Bob's state is the vacuum state and 
Eve can also operate on a vacuum state instead of the intercepted signal. In this scenario Bob and Eve 
have the same mutual information with Alice. This suggests a possible attack on the
present protocol, where for very high losses Eve chooses to operate on the vacuum state instead of her share of the 
intercepted signal. This seems reasonable since the vacuum state is closer to the state with Bob in a very lossy
environment. If Eve chooses to work with both the intercepted signal and 
the vacuum state, the effective secure key rate that can be achieved between Alice and Bob is  
\begin{equation}
K_e = \max\{0,\min\{\Delta I,\Delta I_0\}\}, 
\label{ke}
\end{equation}
where $\Delta I_0 = \beta I_{AB}-I^0_{AE}$ and
$I^0_{AE}$ is the mutual information between Alice and Eve assuming Eve's state 
is the vacuum. 

Using for $g_u$ and $g_v$ their optimal previously obtained expressions when the matching condition occurs, Eqs. (\ref{extra3}) and (\ref{extra4}), 
$K_e$ becomes a function of only $r$, $\theta$, and $\eta$. Fixing $\eta$, we can optimize $K_e$ 
as a function of $r$ and $\theta$ once we choose a coherent state $|\alpha\rangle$ (see Appendix \ref{D}). Working with $\beta=0.8$, we obtained
for $90\%$ loss two regions of $\alpha$ in which a meaningful key can be obtained. For $\alpha \approx 0.5$ we have $K_e = 0.001$
with $r = 1.44$ ($12.5$ dB) and for $\alpha \approx 1.75$ we get $K_e \approx 0.009$ with $r \approx 0.93$ ($8.1$ dB). When
the losses are $95\%$ we get for $\alpha \approx 0.5$, $K_e \approx 0.0008$ with $r \approx 1.47$ ($12.8$ dB), and for
$\alpha \approx 1.6$, $K_e \approx 0.001$ with $r \approx 0.85$ ($7.4$ dB). In Appendix \ref{E} we give for every $\alpha$ between
$0$ and $10$ the optimal values for the key rates and the corresponding optimal parameters leading to those key rates.

\begin{table}[!ht]
\caption{\label{tab1} Optimized key rates $K_e$ for a fixed reconciliation efficiency $\beta$,
loss $1-\eta$, and squeezing $r$ with the corresponding optimal parameters. 
We assume real states $|\alpha\rangle$.}
\begin{center}
\begin{tabular}{cccccccc}
\hline
$\beta$ & loss & $r$(dB) &  $\alpha$ & $\cos^2\theta$& $g_u$ & $g_v$ & $K_e$ \\ \hline\hline
0.8 & 95\% & $0.9(7.82)$ & 1.65 & 0.058 & 0.957 & 0.632 & $2.9\times 10^{-3}$ \\
0.9 & 95\% & $0.7(6.08)$ & 1.60 & 0.080 & 0.887 & 0.494 & $3.7\times 10^{-3}$ \\
1.0 & 99\% & $0.1(0.87)$ & 1.50 & 0.114 & 0.186 & 0.068 & $3.0\times 10^{-6}$ \\
1.0 & 99\% & $0.2(1.74)$ & 1.50 & 0.116 & 0.360 & 0.139 & $6.8\times 10^{-5}$ \\
1.0 & 99\% & $0.3(2.61)$ & 1.50 & 0.108 & 0.516 & 0.206 & $4.2\times 10^{-4}$ \\
1.0 & 99\% & $0.4(3.47)$ & 1.55 & 0.129 & 0.640 & 0.306 & $1.3\times 10^{-3}$  \\
\hline
\end{tabular}
\end{center}
\end{table}

Fortunately the present scheme can be secure for low squeezing $r$, in particular if Alice and Bob use state of the art reconciliation protocols
in which $\beta \approx 1$. Working with fixed values of $r$ and maximizing $K_e$ as a function of $\theta$ we can show that 
for squeezing below $2$ dB it is still possible to get a secure key. In table \ref{tab1} we show the maximum $K_e$ attainable 
for different values of squeezing and reconciliation efficiencies ($\beta$). Note that for $\beta = 0.8$ or $0.9$ our numerical maximization 
indicated that one cannot get a secure key when losses are about or higher than $99\%$ .

We can estimate how far Alice and Bob can be for the present protocol to work securely and with
a reasonable bit generation rate as follows. Noting that state of the art generation rates of two-mode squeezed states \cite{Fur07} are of 
$10^6$ events per second, and working with a key rate of at least $K_e=10^{-3}$, the present protocol allows Alice and Bob to
share at least $10^3$ bits per second. With that in mind, looking at table \ref{tab1} we see that for $\beta = 0.8$ we can get $K_e=10^{-3}$ at
$95\%$ loss and for $\beta = 1.0$ we can go up to $99\%$ loss. Now, assuming standard telecommunication fiber optics,
we have an attenuation coefficient $\epsilon = 0.2 \,\mbox{dB/km}$. Since the relation between distance $L$, loss
$1- \eta$ and $\epsilon$ is $\eta = 10^{-\epsilon L/10}$ \cite{Gis02}, we get for $95\%$ loss $L=65$ km and for
$99\%$ loss $L=100$ km. 

We have also computed for several values of reconciliation efficiency $\beta$ and squeezing $r$ the optimal key rate as
a function of the loss or, equivalently, the distance between Alice and Bob. 
The free parameters in the optimization procedure were the coherent state $\alpha$ 
and Alice's BS transmittance $\cos^2\theta$; $g_u$ and $g_v$ were set to the values given in Eqs. (\ref{extra3}) and (\ref{extra4}).

\begin{figure}[!ht]
\begin{center}
\includegraphics[width=8cm]{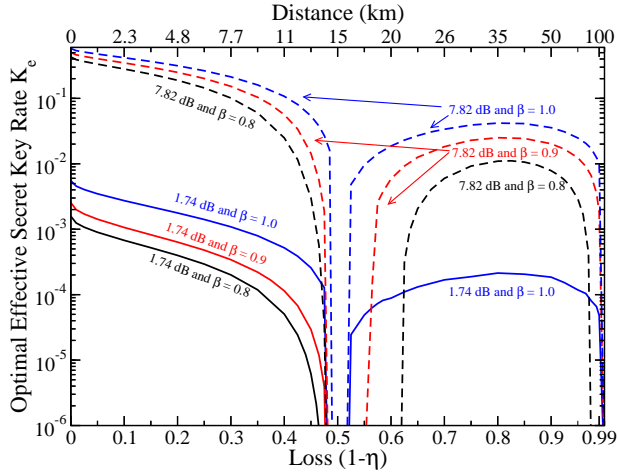}
\end{center}
\caption{\label{keyRateDistance} Optimal key rate as a function of loss (lower horizontal axis) or,
equivalently, as a function of the distance between Alice and Bob (upper horizontal axis), assuming standard 
fiber optics attenuation ($0.2 \,\mbox{dB/km}$). The squeezing $r$ (in dB) of the quantum channel connecting 
Alice and Bob as well as the reconciliation efficiency $\beta$ employed in the computation of the key are indicated in 
the curves.
}
\end{figure}

\begin{figure}[!ht]
\begin{center}
\includegraphics[width=8cm]{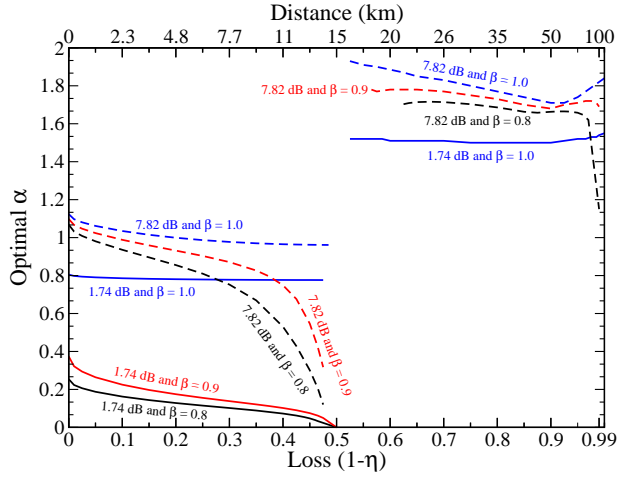}
\end{center}
\caption{\label{keyRateAlfa} Coherent state $\alpha$ leading to the optimal key rates shown in Fig. \ref{keyRateDistance}.
}
\end{figure}

\begin{figure}[!ht]
\begin{center}
\includegraphics[width=8cm]{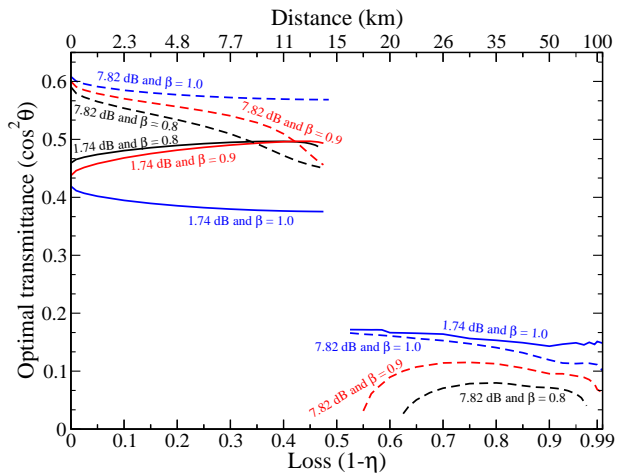}
\end{center}
\caption{\label{keyRateTransmittance} Alice's BS transmittance $\cos^2\theta$ leading to the optimal key rates shown in Fig. \ref{keyRateDistance}.
The optimal $g_u$ and $g_v$ can be obtained using Eqs. (\ref{extra3}) and (\ref{extra4}) and the data above to obtain 
$\theta$ by noting that $0<\theta <\pi/2$.
}
\end{figure}

In Fig. \ref{keyRateDistance} we show the optimal key rate $K_e$, Eq. (\ref{ke}), 
and in Figs. \ref{keyRateAlfa} and \ref{keyRateTransmittance} the optimal parameters leading to the curves shown
in Fig. \ref{keyRateDistance}.

Looking at Fig. \ref{keyRateDistance} we note that there exist two distinct regimes for the behavior of 
the optimal key rate $K_e$. The first one, for losses below $50\%$, as we increase the loss we decrease 
$K_e$. The second regime occurs for losses greater than $50\%$. In this case $K_e$ first increases with more
loss, reaching its maximum value at about $80\%$ loss, and then decreases with loss. It is worth
mentioning that at losses about $50\%$ no key can be established, at least to the precision of our numeric computations ($6$ significant figures). 
We can understand that fact
remembering that at
$50\%$ loss the density matrices describing the states with Bob and
Eve are equal and, therefore, the mutual information between Bob and Alice is exactly equal to the
one between Eve and Alice; no secure key can be established in this case \footnote{In other words, as we approach the value of $50\%$ loss, 
either from above or below, the states reaching Bob and Eve become 
more and more equal and the key rate must necessary decrease, being exactly zero when we reach the $50\%$ loss threshold
since in this situation Bob and Eve have exactly the same state. }. 
For a similar reason we cannot get a secure key
for losses close to $100\%$, since in this situation Eve employs the vacuum state which is very close to the state with Bob, who receives 
almost no signal. Thus, it is expected that very close to the $50\%$ loss or to the $100\%$ loss no key 
can be generated. Moreover, as the loss approaches $50\%$, either from below or above, and as
the loss tends to $100\%$, the key rate decreases very fast, being exactly zero at those two values of
loss for the reasons given above. 

Looking at Fig. \ref{keyRateTransmittance} we can see the main reason why this protocol works 
securely when Eve implements the BS attack and operates on her share of the signal in the \textit{same} way as Bob, 
even when we cross the $50\%$ loss threshold. 
It is due to the fact that the optimal transmittance ($\cos^2\theta$) for Alice's BS leading to the 
highest mutual information between 
Alice and Bob, and therefore to the optimum key rate $K_e$, depends on the loss of the quantum channel connecting them.
(We should not forget that Alice's BS transmittance also depends on whether we have real or imaginary states. We are
considering, as stated in the beginning of this section, 
the situation where Alice employed the real basis and both Bob and Eve correctly chose the real displacements.)
Indeed, since Bob receives $\eta$ of the signal in a lossy transmission, Alice sets the transmittance of her BS 
in order to maximize the mutual information between her 
and Bob in this scenario. However, Eve gets $1-\eta$ of the signal, which requires a completely different value of 
transmittance for Alice's BS in order to make Eve's state a good approximation to the teleported one. In other words, 
since Alice chooses the optimal setting for her BS according to the intensity of the signal reaching Bob ($\eta$), Eve's 
share of the teleported state is not as good a description of the original teleported state as Bob's share is. 
Because of this fact the mutual information between Alice and Eve ($I_{AE}$) is lower than the one between Alice and 
Bob ($I_{AB}$), which is the ingredient needed to establish a secure key between Alice and Bob.  

A final remark is in order before we finish this section. 
The previous security analysis was carried out assuming an individual (incoherent) BS
attack and Eve operating \textit{exactly} as Bob in order to extract the secure key.
Therefore, it is important to extend the security analysis here in at least two ways 
to check whether the interesting security properties of the present protocol still hold, in particular its
secure operation above the $50\%$ loss threshold. First,
we need to check different scenarios at the incoherent attack level. For example,   
what would happen if Eve attenuates her share of the signal to the same intensity reaching Bob and only then operates on her
share to extract the key? Second, it is crucial to study more powerful attacks, 
such as collective and coherent attacks. 
Moreover, it is also important to point out that it is not obvious that the techniques employed in the security analysis of collective and coherent attacks 
for Gaussian modulated CV protocols can be directly employed here (we employed binary discrete modulation/encoding). 
This is due to the fact that a non-Gaussian encoding of the key,
even if employing Gaussian states, have non-Gaussian entanglement-based representations, and the latter fact 
means that the calculation of Eve's information cannot rely on 
the optimality proofs of continuous modulated protocols  \cite{Bra05}.  In other words, a whole new mathematical
analysis must be done in order to compute in our case the optimal Holevo's bound (upper bound of Eve's information), 
the quantity needed to investigate collective 
and more general forms of security attacks. So far we could not solve that problem or find its solution in the literature.
A possible starting point in this direction would be to generalize the analysis in Ref. \cite{Bec13} to the present
protocol in order to estimate at least lower and upper bounds on the secure key rate when Eve implements collective attacks.

We also remark that our main goal in writing this article 
was to present a new way of doing CVQKD based on the CV teleportation protocol 
and to understand its \textit{potential} as a viable secure alternative to realize CVQKD.  
We compared its efficiency and security to the ones of the standard CVQKD protocols when those protocols operate 
under the same assumptions as ours, i.e, a BS attack (loss in the channel) and no excess noise. 
The standard CVQKD protocols we used as a benchmark of comparison were those of Refs. \cite{Gro02,Leu02}.
And when it comes to efficiency in the security scenario described above, our protocol gives higher key rates
than the ones of the aforementioned references. This is the main message we wanted to pass by writing this article and we hope to encourage
those working with CVQKD to further assess the security of the present protocol under more severe attacks.

\section{Conclusion}

In summary, we proposed a new and efficient CVQKD scheme 
with a binary encoding for the key (discrete modulation) based on the CV teleportation of coherent states, where
the CV teleportation protocol is not just a substitute to the direct sending of coherent states from Alice to Bob 
for the usual CVQKD protocols. 
Rather, the resources needed to implement the CV teleportation protocol play a direct role in the generation of the secret key
since Alice's BS transmittance, the squeezing of the entangled channel, and Bob's displacement are all tuned in order
to generate a secret key. 

We showed that the present teleportation-based CVQKD protocol is secure against individual attacks and in particular that it works with
direct reconciliation and no postselection even for very high loss in the optical channel connecting Alice and Bob. 
Moreover, we showed that it is possible to achieve fairly high key rates with mild squeezing ($\approx 2$ dB) near the 
$100\%$ loss regime. This fact combined with the high repetition rates of CV technology may lead to efficient long distance
QKD protocols. Indeed, once a mildly squeezed two-mode entangled state channel is established between Alice and Bob, directly or
via entanglement swapping techniques, they can generate a secret key using the present CVQKD scheme.

Finally, the present CVQKD protocol naturally leads to many interesting open questions. First, since we have only dealt with the
noiseless case, the next question is to understand how robust  
the present scheme is to the addition of noise at the transmission line. Second, can 
reverse reconciliation and/or postselection increase the key rates of this scheme and decrease even more the level 
of squeezing to generate a secure key? Third, will the present protocol still work in a very lossy environment if it
suffers different types of security attacks, such as the collective and coherent attacks? 
Those are the problems we will be tackling in the near future and, so far, the main message one can extract from the
present article is that for individual BS attacks we have a teleportation-based CVQKD protocol, built on a binary encoding
for the key, with at least the same level of security of the standard CVQKD protocols and, at the same time, 
operating beyond the $50\%$ loss threshold without resorting to postselection or reverse reconciliation.

\section*{Acknowledgments}
FSL and GR thank CNPq (Brazilian National Council for Scientific and Technological Development)
for funding and GR thanks CNPq/FAPESP (State of S\~ao Paulo Research Foundation)
for financial support through the National Institute of
Science and Technology for Quantum Information.

\appendix

\section{The modified CV teleportation protocol}
\label{B}

A key ingredient to the present scheme is the CV teleportation protocol \cite{Bra98} adapted to the case where
Alice and Bob has a complete knowledge of the pool of possible states to be teleported \cite{Lui14}. 
With such a knowledge, Alice and Bob can greatly improve the fidelity between the teleported 
state with Bob and Alice's input by changing certain parameters of the original proposal.
Our goal in this section is to review in a self contained way
this modified CV teleportation protocol, following closely the presentation given in \cite{Lui14}.

Let $\hat{x}_k=(\hat{a}_k+\hat{a}_k^\dagger)/2$ and $\hat{p}_k=(\hat{a}_k-\hat{a}_k^\dagger)/2i$
be the position and momentum quadratures of mode $k$, respectively, where $\hat{a}_k$
and $\hat{a}_k^\dagger$ are the annihilation and creation operators   
with commutation relation $[\hat{a}_k,\hat{a}_k^\dagger]=1$.  

Any input state with Alice can be expressed in the position basis as 
\begin{equation}
|\varphi\rangle= \int dx_1\varphi(x_1)|x_1\rangle,
\label{input}
\end{equation}
where the integral covers the entire real line and $\varphi(x_1)= \langle x_1 | \varphi \rangle$. 
The entangled two-mode squeezed state shared between Alice and Bob can 
also be expressed in the position basis,
\begin{equation}
|\psi_r\rangle = \int dx_2 dx_3 \psi_r(x_2, x_3) |x_2, x_3 \rangle, 
\label{canal}
\end{equation}
with $\psi_r(x_2,x_3)= \langle x_2, x_3 | \psi_r \rangle$ and 
$|x_2, x_3 \rangle = |x_2\rangle \otimes |x_3 \rangle$.  Here 
the first two modes/kets are with Alice and the third one with Bob. Using 
Eqs.~(\ref{input}) and (\ref{canal}) the initial state describing all
modes before the teleportation is as follows,
\begin{eqnarray}
|\Psi \rangle &=& |\varphi\rangle \otimes |\psi_r\rangle =\int dx_1 dx_2 dx_3 
\varphi(x_1) \psi_r(x_2,x_3)|x_1,x_2,x_3\rangle.
\label{initial}
\end{eqnarray}

The teleportation begins sending mode 1 (input state) and mode 2 (Alice's
share of the entangled state) into a BS with transmittance $\cos^2\theta$
(see figure \ref{fig_esquema}). If $\hat{B}_{12}(\cos^2\theta)$ is the operator representing the action 
of the BS in the position basis we have \cite{Bra05}
\begin{equation}
\hat{B}_{12}(\cos^2\theta) |x_1, x_2 \rangle = |x_1 \sin\theta + 
x_2 \cos\theta, x_1 \cos\theta - x_2 \sin\theta\rangle.
\label{bs}
\end{equation}

Inserting equation (\ref{bs}) into (\ref{initial}) and changing variables such that 
$x_v = x_1 \sin\theta + x_2 \cos\theta$ and 
$x_u =x_1 \cos\theta - x_2 \sin\theta$ we get
\begin{eqnarray}
|\Psi' \rangle &=&\int dx_v dx_u dx_3 
\varphi(x_v\sin\theta + x_u\cos\theta)\nonumber \\
&& \times \psi_r(x_v\cos\theta - x_u\sin\theta,x_3)|x_v,x_u,x_3\rangle
\label{initial2}
\end{eqnarray}
for the total state after modes $1$ and $2$ go through the BS.

In the next step Alice measures the momentum and position quadratures
of modes $v$ and $u$, respectively.  
Since Alice will project mode $v$ onto the momentum basis, it is convenient to rewrite
equation (\ref{initial2}) using the Fourier transformation 
relating the position and momentum basis, 
\begin{equation}
|x_v \rangle = \frac{1}{\sqrt{\pi}} \int dp_v 
e^{- 2i x_v p_v} |p_v\rangle.
\label{fourier1}
\end{equation}
This leads to 
\begin{eqnarray}
|\Psi' \rangle &=&\frac{1}{\sqrt{\pi}}\int dp_v dx_v dx_u dx_3 \varphi(x_v\sin\theta + x_u\cos\theta)
\nonumber \\
&&\times \psi_r(x_v\cos\theta - x_u\sin\theta,x_3)e^{- 2i x_v p_v}|p_v,x_u,x_3\rangle.
\nonumber \\
\label{initial2b}
\end{eqnarray}

Let us assume Alice obtains for the momentum of mode $v$ and for the position of
mode $u$ the values $\tilde{p}_v$ and $\tilde{x}_u$.
Thus, the state after the measurement is 
$$
|\Psi''\rangle=\hat{P}_{\tilde{p}_v,\tilde{x}_u}|\Psi'\rangle/
\sqrt{\mathbb{p}(\tilde{p}_v,\tilde{x}_u)},
$$ 
where 
$\hat{P}_{\tilde{p}_v,\tilde{x}_u}=
|\tilde{p}_v,\tilde{x}_u\rangle\langle\tilde{p}_v,\tilde{x}_u| \otimes \mathbb{1}_{3}$
is the von Neumann projector describing the measurements. Here
$\mathbb{1}_3$ is the identity operator acting on mode $3$ and 
$\mathbb{p}(\tilde{p}_v,\tilde{x}_u)=\tr(|\Psi'\rangle\langle\Psi'|\hat{P}_{\tilde{p}_v,\tilde{x}_u})$
is the probability of measuring momentum $\tilde{p}_v$ and
position $\tilde{x}_u$, with $\tr$ denoting the total trace. 
Specifying to the position basis and using that 
$\langle p_v|\tilde{p}_v\rangle=\delta(p_v-\tilde{p}_v)$ and
$\langle x_u|\tilde{x}_u\rangle=\delta(x_u-\tilde{x}_u)$ we have
\begin{equation}
|\Psi'' \rangle = |\tilde{p}_v,\tilde{x}_u\rangle \otimes |\chi'\rangle,
\label{quasebob}
\end{equation}
where Bob's state is 
\begin{eqnarray}
|\chi' \rangle&=&\frac{1}{\sqrt{\pi \mathbb{p}(\tilde{p}_v,\tilde{x}_u)}}\int dx_v dx_3 e^{- 2i x_v \tilde{p}_v}
\varphi(x_v\sin\theta + \tilde{x}_u\cos\theta) \nonumber \\
&&\times \psi_r(x_v\cos\theta - \tilde{x}_u\sin\theta,x_3)
|x_3\rangle.
\label{initial3}
\end{eqnarray}
Here 
\begin{equation}
\mathbb{p}(\tilde{p}_v,\tilde{x}_u) = \int dx_3 |\Psi'(\tilde{p}_v,\tilde{x}_u,x_3)|^2
\label{prob}
\end{equation}
and $\Psi'(\tilde{p}_v,\tilde{x}_u,x_3)=\langle \tilde{p}_v,\tilde{x}_u,x_3| \Psi'\rangle$ such that
\begin{eqnarray}
\Psi'(\tilde{p}_v,\tilde{x}_u,x_3)&=&\frac{1}{\sqrt{\pi}}\int dx_v
\varphi(x_v\sin\theta + \tilde{x}_u\cos\theta)  \nonumber \\
&&\times \psi_r(x_v\cos\theta - \tilde{x}_u\sin\theta,x_3)e^{- 2i x_v \tilde{p}_v},
\label{psilinha}
\end{eqnarray}
where equation (\ref{psilinha}) was obtained using (\ref{initial2b}).

Via a classical channel Alice sends to Bob her measurement results, allowing 
Bob to displace his mode quadratures as follows,
$x_3 \rightarrow x_3 + g_u\tilde{x}_u$ and $p_3 \rightarrow p_3 + g_v\tilde{p}_v$. 
Mathematically this corresponds to the application of the displacement operator 
$\hat{D}(\lambda)=e^{\lambda \hat{a}^\dagger-\lambda^*\hat{a}}=
e^{-2iRe[\lambda]\hat{p}+2iIm[\lambda]\hat{x}}$, with 
$\lambda = g_u\tilde{x}_u + ig_v\tilde{p}_v$ and $\,^*$ denoting the complex conjugation. 
Since $\hat{x}$ and $\hat{p}$ commute with
their commutator Glauber's formula applies, giving 
$\hat{D}(\lambda)=e^{iRe[\lambda]Im[\lambda]}e^{-2iRe[\lambda]\hat{p}}e^{2iIm[\lambda]\hat{x}}$
and finally 
\begin{equation}
\hat{D}(g_u\tilde{x}_u + ig_v\tilde{p}_v)|x_3\rangle = e^{ig_ug_v\tilde{x}_u\tilde{p}_v}e^{2ig_v\tilde{p}_vx_3}
|x_3 + g_u\tilde{x}_u\rangle.
\label{disp}
\end{equation}

Bob's state after the displacement, $|\chi\rangle=\hat{D}(g_u\tilde{x}_u + ig_v\tilde{p}_v)|\chi'\rangle$, can be written as follows if
we use equation (\ref{disp}) and change variable such that
$x_3 \rightarrow x_3 - g_u\tilde{x}_u$,
%
\begin{eqnarray}
|\chi \rangle =\int dx_3 \chi(x_3) |x_3\rangle,
\label{bobfinal}
\end{eqnarray}
with
\begin{eqnarray}
\chi(x_3)&=&\frac{e^{-ig_ug_v\tilde{x}_u\tilde{p}_v}}{\sqrt{\pi \mathbb{p}(\tilde{p}_v,\tilde{x}_u)}}
\int dx_v\varphi(x_v\sin\theta + \tilde{x}_u\cos\theta) \nonumber \\
&&\times \psi_r(x_v\cos\theta - \tilde{x}_u\sin\theta,x_3-g_u\tilde{x}_u)
e^{- 2i (x_v -g_vx_3)\tilde{p}_v}. 
\end{eqnarray}

In order to estimate after a single run of the protocol the closeness of Bob's state, $\hat{\rho}_B=|\chi \rangle\langle\chi|$,
with the original one at Alice's, $\rho_{input}=|\varphi\rangle\langle \varphi|$, we use the fidelity
\begin{eqnarray}
F=\langle \varphi | \hat{\rho}_B | \varphi \rangle = \int dx_3'dx_3 \varphi^*(x_3')\chi(x_3')\chi^*(x_3)\varphi(x_3).
\label{fid1}
\end{eqnarray}
In general $F$ depends on the input state $|\varphi\rangle$,
the measurement outcomes of Alice ($\tilde{x}_u$ and $\tilde{p}_v$), the squeezing $r$ of the 
entangled two-mode squeezed state, $\theta$, $g_u$, and $g_v$. 
Also, F achieves its highest value ($F=1$) if we
have a flawless teleportation ($\hat{\rho}_B = \hat{\rho}_{input}$) and its minimal one ($F=0$)
if the output is orthogonal to the input.

We will be dealing with input states given by coherent states,
$|\varphi\rangle=|\alpha e^{i\xi}\rangle$, with $\alpha$ and $\xi$ reals, and with entangled two-mode squeezed states shared between
Alice and Bob $|\psi_r\rangle = \sqrt{1-\tanh^2r}\sum_{n=0}^{\infty}\tanh^nr|n\rangle_A\otimes|n\rangle_B$,
where $|n\rangle_{A(B)}$ are Fock number states with Alice (Bob) and $r$ is the squeezing
parameter. When $r=0$ we have $|00\rangle$, the vacuum state, and for $r\rightarrow \infty$ the
unphysical maximally entangled Einstein-Podolsky-Rosen (EPR) state. 

In the position basis we have \cite{Bra05}
\begin{equation}
\varphi(x_1)=\langle x_1 | \alpha e^{i\xi} \rangle = \left(\frac{2}{\pi}\right)^{1/4}e^{-x_1^2+2\alpha e^{i\xi} x_1
-\alpha^2/2-\alpha^2e^{i2\xi}/2}
\label{in}
\end{equation}
and
\begin{equation}
\hspace{-2cm}\psi_r(x_2,x_3)= \langle x_2,x_3 | \psi_r \rangle =
\sqrt{\frac{2}{\pi}}\exp\left[-e^{-2r}(x_2+x_3)^2/2\right.\left.-e^{2r}(x_2-x_3)^2/2\right].
\label{ch}
\end{equation}

Note that for a two-mode squeezed state the variance 
$\Delta_r^2(x_2-x_3)=\langle \psi_r| (x_2-x_3)^2 |\psi_r \rangle-\langle \psi_r| (x_2-x_3) |\psi_r \rangle^2=e^{-2r}/2$,
which is employed to measure the squeezing of this state in decibel: 
\begin{equation}
I_{\mbox{dB}} = -10 \log_{10}\left[\frac{\Delta_r^2(x_2-x_3)}{\Delta_0^2(x_2-x_3)}\right]=20r\log_{10}(e).
\end{equation}

\section{The teleportation-based CVQKD protocol}
\label{C}

The present CVQKD protocol is based on a binary encoding for the key such that 
$\{|-\alpha\rangle,|\alpha\rangle,|-i\alpha\rangle,|i\alpha\rangle\}=\{0,1,0,1\}$, with $\alpha$ a real
number. These states are to be teleported from Alice to Bob randomly. 
A step by step description of a successful run of the protocol, generating a common random bit between
Alice and Bob, is as follows. 
(1) Alice randomly chooses between the real or imaginary coherent state ``basis'' and then randomly prepares 
$|\pm\alpha\rangle$ or $|\pm i\alpha\rangle$, respectively, to teleport to Bob. In Fig. \ref{fig_esquema} we describe the case
where Alice chooses $|\alpha\rangle$ (mode $1$ given by the solid/blue line). (2) Alice generates a two-mode squeezed
entangled state (modes $2$ and $3$), whose squeezing parameter $r$ is chosen according to the value of
$\alpha$, and sends mode $3$ to Bob. (3) Alice adjusts the beam splitter (BS)
transmittance according to her choosing the real or imaginary basis and then sends mode $1$ to interact with her
share of the two-mode squeezed state (mode $2$). (4) She measures the position and momentum quadratures of the modes $u$ and $v$, respectively, that emerge
after the BS and classically informs Bob of those results ($\tilde{x}_u$ and $\tilde{p}_v$). 
(5) Bob randomly chooses $(g_u, g_v)$ from two possible pairs of values and implements a displacement operation on his mode given by 
$\hat{D}(\lambda)$, where $\lambda = g_u\tilde{x}_u+ig_v\tilde{p}_v$.
$g_u$ and $g_v$ are such that the fidelity of Bob's output state with Alice's input is greatest if she chooses a real (imaginary) state and 
he assumes a real (imaginary) state and, at the same time, least if she chooses an imaginary (real) state and he assumes a real (imaginary) state. 
The optimal pair ($g_u$, $g_v$) depends on the input being a real or imaginary coherent state but not on its sign. (6) Bob implements another
displacement on his mode, $\hat{D}(\alpha)$ or $\hat{D}(i\alpha)$, depending on the choice he made for the pair  $(g_u, g_v)$.
Fig. \ref{fig_esquema} shows the case in which Bob assumes Alice chooses the real basis (solid lines). Had he assumed the wrong basis, which Alice and Bob will
discover classically communicating after finishing the whole protocol, they would discard this run of the protocol.
(7) Bob measures the intensity of his mode and assigns the bit value $0$ if he sees no light (vacuum mode) and the bit $1$ otherwise.

\subsection{Fidelity analysis}

We will explicitly analyze the case where Alice chooses the real basis, namely, she teleports either $|-\alpha\rangle$ 
or $|\alpha\rangle$ to Bob. The calculations for the imaginary basis are similar and only the final results for this case will
be given. Therefore, assuming that we have a real coherent state, Eqs. (\ref{bobfinal}), (\ref{in}), and (\ref{ch}) when inserted into Eq. (\ref{fid1})
give
%
\begin{equation}
\hspace{-1cm} F = h_1(r,\theta)\exp[f_1(\tilde{p}_v,\tilde{x}_u,g_v,g_u,r,\theta)+2\alpha \tilde{x}_u f_2(g_u,r,\theta)+\alpha^2 f_3(r,\theta)],
\label{fidr}
\end{equation}
where 
\begin{eqnarray*}
h_1(r,\theta)&=&\sqrt{1-\cos ^2(2 \theta )\tanh ^4r },\\
f_1(\tilde{p}_v,\tilde{x}_u,g_v,g_u,r,\theta) &=&  \left\{ \left[g_u^2 \tilde{x}_u^2-g_v^2 \tilde{p}_v^2\right]\cos (2 \theta ) \tanh r +4 g_u \tilde{x}_u^2
   \sin\theta \right.\\
   && \left. +4 g_v \tilde{p}_v^2 \cos \theta \right\}\tanh r \\
   &&+\tilde{x}_u^2 \left[-g_u^2+\frac{2}{\cosh ^2r- \cos (2 \theta )\sinh ^2r}-2\right] \\ 
    && +\tilde{p}_v^2 \left[-g_v^2+\frac{2}{\cosh^2r+\cos (2 \theta )\sinh^2r }-2\right], \\
f_2(g_u,r,\theta) &\hspace{-.1cm}=&\hspace{-.1cm}
g_u -\{g_u  \cos (2 \theta )\tanh r+2  [1+g_u \cos \theta]\sin \theta\}\tanh r \nonumber \\
&&+2 \left(\frac{1}{ \cos (2 \theta )\sinh ^2r-\cosh ^2r}+1\right)\cos \theta, \\
f_3(r,\theta)&=& -\frac{\{\cosh r- [\tanh r \cos (2 \theta )+\sin (2 \theta )]\sinh r\}^2}{\cosh ^2r- \cos (2 \theta )\sinh ^2r}. 
\end{eqnarray*}

Since we want the optimal $F$ in a way that the optimal settings do not depend on the sign of $\alpha$ we set $f_2(g_u,r,\theta)=0$. This
gives the following value for $g_u$,
\begin{equation}
g_u(r,\theta) = \frac{\sinh (2 r) \sin \theta}{\cosh^2r-\cos(2 \theta )\sinh^2r }.
\label{gur}
\end{equation}

Moreover, since $g_v$ only appears in the exponent and we want the maximum of $F$, we maximize the exponent as a function of $g_v$. Differentiating
the exponent with respect to $g_v$ and equating to zero we get
\begin{equation}
g_v(r,\theta) = \frac{2 \coth r \cos\theta}{\coth^2r+\cos(2 \theta )}.
\label{gvr}
\end{equation}

Inserting $g_u$ and $g_v$ back into $F$ we finally obtain 
\begin{eqnarray}
F^{\mbox{re}}(r,\theta) &=& \sqrt{1- \cos^2(2 \theta)\tanh^4r}  \nonumber \\
&\times& \exp\left\{\frac{-\alpha^2 \{\cosh r-\sinh r [ \cos (2 \theta )\tanh r+
\sin (2\theta )]\}^2}{\cosh^2r-\cos(2 \theta )\sinh^2r }\right\}, \nonumber \\
&&\,
\label{fidre}
\end{eqnarray}
where we use the superscript ``re'' to remind us that this is the optimal $F$ for real inputs.
Also, it is important to note that the optimal expression for $F$, as well as for $g_u$ and $g_v$, do not depend
on the measurement outcomes $\tilde{x}_u$ and $\tilde{p}_v$ obtained by Alice. This is one of the reasons making 
the present CVQKD scheme yield high key rates without postselecting a subset of all possible measurement outcomes of Alice. 

For an imaginary input, namely, either $|i\alpha\rangle$ or $|-i\alpha\rangle$, the roles of $g_u$ and $g_v$ are reversed. In order to have a solution for $F$ independent of the sign
of the imaginary coherent state we fix $g_v$. Then, we maximize the exponent of $F$ as a function of $g_u$. The final result is that
we obtain the same expressions for $g_u$ and $g_v$ as given before for the real case and the following expression for the  fidelity:
\begin{eqnarray}
F^{\mbox{im}}(r,\theta) &=& \sqrt{1- \cos^2(2 \theta)\tanh^4r} \nonumber \\
&\times & \exp\left\{\frac{-\alpha^2 \{\cosh r+\sinh r [ \cos (2 \theta )\tanh r-
\sin (2\theta )]\}^2}{\cosh^2r+ \cos(2 \theta )\sinh^2r}\right\}. \nonumber \\
&& \,
\label{fidim}
\end{eqnarray}

Comparing both expressions for the fidelity we see that 
\begin{equation}
F^{\mbox{re}}(r,\theta) = F^{\mbox{im}}(r,\pi/2-\theta).
\end{equation}

The final calculations needed to determine the optimal $r$ and $\theta$ are as follows. We want 
$r$ and $\theta$ such that if Alice chooses the real basis and Bob assumes Alice chose the real basis, 
$F^{\mbox{re}}$ is maximal and $F^{\mbox{im}}$ is minimal. This is achieved maximizing the following function:
\begin{equation}
\Pi^{\mbox{re}}(r,\theta) = F^{\mbox{re}}(r,\theta) [1-F^{\mbox{im}}(r,\theta)].
\label{pire}
\end{equation}

It is not possible, however, to analytically solve the optimization problem associated to Eq.~(\ref{pire}) and
get simple closed expressions for the optimal $r$ and $\theta$. Thus, the maximization of Eq.~(\ref{pire}) 
is carried out numerically once the value of $\alpha$ is specified. This is what was done to get the optimal data shown
in figure 2 of the main text.

The optimal parameters if Alice chooses the imaginary basis and Bob assumes Alice chose the imaginary basis is obtained 
imposing that $F^{\mbox{re}}$ be minimal and $F^{\mbox{im}}$ be maximal. This is obtained maximizing the following function:
\begin{equation}
\Pi^{\mbox{im}}(r,\theta) = F^{\mbox{im}}(r,\theta) [1-F^{\mbox{re}}(r,\theta)]=\Pi^{\mbox{re}}(r,\pi/2-\theta).
\end{equation}
It is clear by the last equality that the optimal $\theta$ for the imaginary input is obtained from the optimal one for the
real input by subtracting it from $\pi/2$. The relations between the optimal settings for the real and imaginary inputs are as follows:
\begin{eqnarray}
\theta^{\mbox{re}} &=& \pi/2 - \theta^{\mbox{im}}, \\
g_v^{\mbox{re}} &=& g_u^{\mbox{im}}, \\
g_u^{\mbox{re}} &=& g_v^{\mbox{im}}, \\
r^{\mbox{re}} &=& r^{\mbox{im}}.
\end{eqnarray}

\subsection{Key generation analysis}

The state with Bob after finishing the teleportation protocol is given by equation (\ref{bobfinal}), where he has already
implemented either the real or imaginary displacement on his mode.  By real and imaginary displacements
we mean that Bob applied the displacement $\hat{D}(\lambda)$, with $\lambda = g_u\tilde{x}_u+ig_v\tilde{p}_v$, using either
the real ($g_u^{\mbox{re}}$ and  $g_v^{\mbox{re}}$) or imaginary ($g_u^{\mbox{im}}$ and  $g_v^{\mbox{im}}$)
optimal parameters.

In the next step of the teleportation-based CVQKD protocol, he implements another displacement, which depends on whether he chose the real or imaginary displacement.
For a previously real displaced mode he now applies the displacement $\hat{D}(\alpha)$ and for a previously imaginary displaced mode he
applies $\hat{D}(i\alpha)$. The goal of these last displacements is to transform states nearly described by $|-\alpha\rangle$ or $|-i\alpha\rangle$
to vacuum states and to push further away from the vacuum the states $|\alpha\rangle$ or $|i\alpha\rangle$. Note that Bob's state will be very close to one
of those four states only if the ``matching condition'' occurred, i.e., if Alice teleported a real (imaginary) state and Bob used the optimal settings presuming a
real (imaginary) input by Alice. 

Mathematically, the state after the last displacement is
\begin{equation}
|\tilde{\chi}\rangle = \hat{D}(\gamma)|\chi\rangle, 
\end{equation}
where $\gamma = \alpha$ or $\gamma = i\alpha$. The probability to detect the vacuum state is
\begin{equation}
Q^B_0 = |\langle 0|\tilde{\chi}\rangle|^2=|\langle -\gamma|\chi\rangle|^2
= \left|\int dx_3\varphi^*_{{-\gamma}}(x_3)\chi(x_3)\right|^2\hspace{-.1cm},
\label{Q0}
\end{equation}
where we used that $\hat{D}(\gamma)=\hat{D}^\dagger(-\gamma)$ and $\langle 0|\hat{D}(\gamma)=\langle -\gamma|$.
In equation (\ref{Q0}) $\varphi^*_{{-\gamma}}(x_3)$ is the complex conjugate of (\ref{in}), with the subscript $-\gamma$
as a reminder to which coherent state the kernel $\varphi(x_3)$ refers to, and  $\chi(x_3)$ is given by
equation (\ref{bobfinal}).

\begin{figure}[!ht]
\begin{center}
\includegraphics[width=8cm]{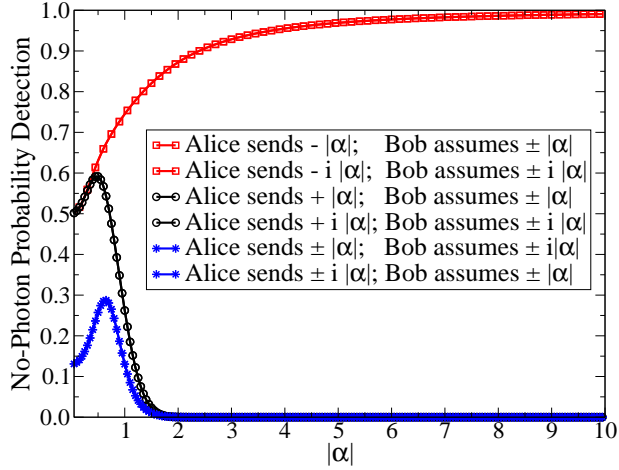}
\end{center}
\caption{ \label{Fig3Main}
The first curve is $Q^B_0$ computed with the following parameters, 
$\mbox{Alice's input} = |-\alpha\rangle,r^{\mbox{re}},\theta^{\mbox{re}},\lambda=g_u^{\mbox{re}}\tilde{x}_u+ig_v^{\mbox{re}}\tilde{p}_v,\gamma=\alpha$.
The second curve is $Q^B_0$ for $\mbox{Alice's input} = |-i\alpha\rangle,r^{\mbox{im}},\theta^{\mbox{im}},\lambda=g_u^{\mbox{im}}\tilde{x}_u+ig_v^{\mbox{im}}\tilde{p}_v,\gamma=i\alpha$.
The third curve is $Q^B_0$ for $\mbox{Alice's input} = |\alpha\rangle,r^{\mbox{re}},\theta^{\mbox{re}},\lambda=g_u^{\mbox{re}}\tilde{x}_u+ig_v^{\mbox{re}}\tilde{p}_v,\gamma=\alpha$.
The fourth curve is $Q^B_0$ for $\mbox{Alice's input} = |i\alpha\rangle,r^{\mbox{im}},\theta^{\mbox{im}},\lambda=g_u^{\mbox{im}}\tilde{x}_u+ig_v^{\mbox{im}}\tilde{p}_v,\gamma=i\alpha$.
The fifth curve is the averaged $Q^B_0$ for $\mbox{Alice's input} = |\pm\alpha\rangle,r^{\mbox{re}},\theta^{\mbox{re}},\lambda=g_u^{\mbox{im}}\tilde{x}_u+ig_v^{\mbox{im}}\tilde{p}_v,\gamma=i\alpha$.
The sixth curve is the averaged $Q^B_0$ for $\mbox{Alice's input} = |\pm i\alpha\rangle,r^{\mbox{im}},\theta^{\mbox{im}},\lambda=g_u^{\mbox{re}}\tilde{x}_u+ig_v^{\mbox{re}}\tilde{p}_v,\gamma=\alpha$.
}
\end{figure}

Figure 3 in the main text is a plot of $Q^B_0$ for all possible combinations of input state by Alice and displacement
by Bob when a matching condition occurs (the first four curves from top to bottom). The fifth and sixth curves 
are $Q^B_0$ averaged over all possible measurement outcomes $\tilde{x}_u$ and $\tilde{p}_v$ for Alice, weighted by Alice's probability 
to get  $\tilde{x}_u$ and $\tilde{p}_v$ (cf. equation (\ref{prob})),
\begin{equation}
q^B_0=\int d\tilde{p}_vd\tilde{x}_u \mathbb{p}(\tilde{p}_v,\tilde{x}_u) Q^B_0(\tilde{p}_v,\tilde{x}_u).
\end{equation}
This averaging is needed whenever the matching condition does not occur since $Q^B_0$ depends on $\tilde{x}_u$ and $\tilde{p}_v$
in this case.
See figure \ref{Fig3Main} for a reproduction of figure 3 of the main text but this time with a different caption, 
where we employ the notation just developed to describe
each one of the plotted curves.

\begin{figure}[!ht]
\begin{center}
\includegraphics[width=8cm]{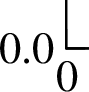}
\end{center}
\caption{ \label{FigNoise} 
For each value of $\alpha$ we have implemented $100$ realizations of random fluctuations about the input state, about the optimal values $r,\theta,g_v,g_u$,
and about $\gamma$.
We worked with Alice's sending a real state and Bob assuming a real state. Similar results are obtained for the imaginary matching condition.
The red/square curves connects the maximal and minimal values for $q^B_0$ due to the random fluctuations assuming Alice sent a negative real state. 
The gray dots between the red/square curves represent the value of $q^B_0$ at each realization. The black/circle curves has the same meaning of the red/square
curves but assuming Alice sent a positive real state.
}
\end{figure}

We have also tested the robustness of the optimal settings by randomly and independently changing the optimal parameters about their correct values.
As can be seen in figure \ref{FigNoise}, the optimal settings are very robust, supporting fluctuations of $\pm 2\%$ about the optimal values for 
small and large $\alpha$. For small $\alpha$ fluctuations of $\pm 10\%$ is still tolerable.

\section{Security analysis}
\label{D}

We want to study how the teleportation-based CVQKD protocol responds to a lossy channel, or equivalently, to the BS attack. 
This will allow us to determine the level of loss in which a secure key can be extracted via direct reconciliation and no postselection.

\subsection{Lossy channel or the presence of Eve}

We want to investigate the security of the present scheme to the BS attack. In the BS attack an eavesdropper (Eve) inserts
a BS of transmittance $\eta$, $0\leq \eta \leq 1$,
during the transmission to Bob of his share of the entangled two-mode squeezed state (mode 3 in figure \ref{fig_esquema}).
In this case Bob will receive a signal with intensity $\eta$ and Eve the rest. With her share of the signal, $1-\eta$, Eve proceeds as Bob in order to extract
information of the key.  

The BS is inserted before Bob receives his mode and therefore before he applies the displacements $\hat{D}(\lambda)$ and $\hat{D}(\gamma)$,
with $\gamma = \alpha$ or $i\alpha$. Bob's state before the insertion of the BS is $|\chi'\rangle$ as given in equation (\ref{initial3}). Hence,
the joint state of Bob and Eve before the BS is
\begin{equation}
|\Omega\rangle = |\chi'\rangle|0\rangle = \int dx_3dx_4 \langle x_3|\chi'\rangle \langle x_4|0\rangle |x_3\rangle|x_4\rangle 
= \int dx_3dx_4 \chi'(x_3) \varphi_0(x_4) |x_3,x_4\rangle,
\end{equation}
with $\varphi_0(x_4)$ given by Eq. (\ref{in}) with $\alpha =0$. But since
\begin{equation}
\hat{B}_{34}(\eta) |x_3, x_4 \rangle = \left| \sqrt{\eta}x_3 -\sqrt{1-\eta}x_4, \sqrt{1-\eta}x_3 + 
 \sqrt{\eta}x_4\right\rangle
\end{equation}
we have after the BS,
%
\begin{eqnarray}
|\Omega\rangle &=&\hat{B}_{34}(\eta)|\chi'\rangle_3|0\rangle_4 \nonumber \\
&=&\int dx_3dx_4 \chi'(x_3) \varphi_0(x_4)  \left| \sqrt{\eta}x_3 -\sqrt{1-\eta}x_4, \sqrt{1-\eta}x_3 + 
 \sqrt{\eta}x_4\right\rangle, \nonumber \\
&=&\!\!\!\int dx_3dx_4 \chi'(\sqrt{\eta}x_3+\sqrt{1-\eta}x_4) \varphi_0(\sqrt{\eta}x_4-\sqrt{1-\eta}x_3) |x_3,x_4\rangle.\nonumber \\
\end{eqnarray}
The last equality was obtained making the following change of variables, $x_3 \rightarrow \sqrt{\eta}x_3+\sqrt{1-\eta}x_4$ and
$x_4 \rightarrow \sqrt{\eta}x_4-\sqrt{1-\eta}x_3$. Bob's state after the BS is given by the partial trace of the state $\rho_{BE}=|\Omega\rangle\langle\Omega|$
with respect to Eve's mode, $\rho'_B=\tr\!\!_E(\rho_{BE})$. In the position basis we have
\begin{equation}
\rho'_B=\int dx_4\langle x_4 | \rho_{BE} |x_4\rangle =\int dx_3dx'_3 \rho'_B(x_3,x'_3)|x_3\rangle\langle x'_3|
\label{rhoBobLoss}
\end{equation}
where
\begin{eqnarray}
\rho'_B(x_3,x'_3)&=&\int dx_4  \chi'(\sqrt{\eta}x_3+\sqrt{1-\eta}x_4) \chi'^*(\sqrt{\eta}x'_3+\sqrt{1-\eta}x_4) \nonumber \\
&&\times \varphi_0(\sqrt{\eta}x_4-\sqrt{1-\eta}x_3)\varphi_0^*(\sqrt{\eta}x_4-\sqrt{1-\eta}x'_3). 
\label{rhoBobLossKernel}
\end{eqnarray}
Note that Eve's state is $\rho'_E=\tr\!\!_B(\rho_{BE})$, which is simply obtained from equation (\ref{rhoBobLossKernel}) by changing $\eta \rightarrow 1-\eta$.

\begin{figure}[!ht]
\begin{center}
\includegraphics[width=8cm]{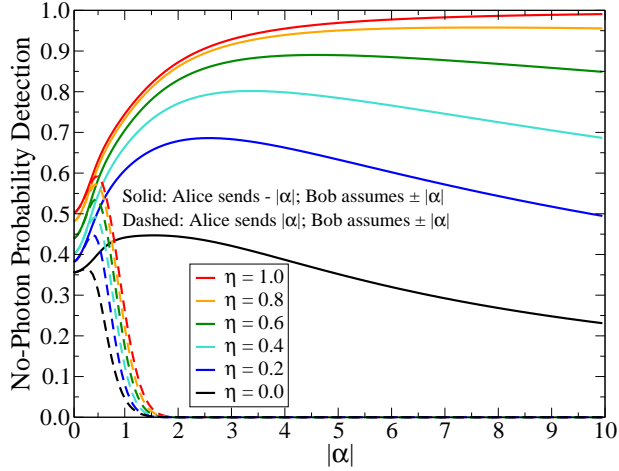}
\end{center}
\caption{ \label{FigBS} 
Probability $q^B_0$ to detect the vacuum state for several values of loss ($1-\eta$),
which increases ($\eta$ decreases) from top to bottom. The other parameters used to compute
$q^B_0$, namely, $r, \theta, g_v, g_u$, and $\gamma$, were the optimal ones when the matching condition occurs.
The remaining parameter, Alice's input, was set to $-\alpha$ (solid lines) and $\alpha$ (dashed lines). 
}
\end{figure}

Using the state $\rho'_B$ ($\rho'_E$) Bob (Eve) proceeds as explained before to finish all the steps of a single run of the teleportation-based CVQKD protocol.
Bob displaces his mode by $\lambda$, which depends on whether he assumed Alice teleported a real or imaginary state, finishing the teleportation stage
of the protocol. His state at this stage is $\rho_B=\hat{D}(\lambda)\rho'_B\hat{D}^\dagger(\lambda)$. Then he implements the last
displacement $\hat{D}(\gamma)$, which depends on his first displacement as explained before, and measures the intensity of his mode. 
Hence, Bob's probability to detect the vacuum state (no-light) is 
\begin{eqnarray}
Q^B_0(\tilde{p}_v,\tilde{x}_u)=\tr[|0\rangle\langle0|\, \hat{D}(\gamma)\rho_B\hat{D}^\dagger(\gamma)] 
= \langle -\lambda -\gamma| \rho'_B|-\lambda -\gamma\rangle, 
\end{eqnarray}
where we have made explicit that $Q^B_0$ depends on the measurement outcomes of Alice when $\eta\neq 1$, i.e., when we have a lossy channel.
In the position representation we have
\begin{equation}
Q^B_0(\tilde{p}_v,\tilde{x}_u)= \int dx_3dx'_3\varphi^*_{\,_{-\lambda -\gamma}}(x_3) \rho'_B(x_3,x'_3)\varphi_{\,_{-\lambda -\gamma}}(x'_3). 
\end{equation}
As before, we define the unconditional (no postselection) probability as
\begin{equation}
q^B_0=\int d\tilde{p}_vd\tilde{x}_u \mathbb{p}(\tilde{p}_v,\tilde{x}_u) Q^B_0(\tilde{p}_v,\tilde{x}_u)
\label{q0}
\end{equation}
and in figure \ref{FigBS} we show its value for several values of loss.

\subsection{Secure key rates}

For direct reconciliation the secure key rate between Alice and Bob is 
\begin{equation}
K=\max\{0,\beta I_{AB} - I_{AE}\}=\{0,\Delta I\},
\label{k}
\end{equation}
where $\beta$ is the reconciliation efficiency, $I_{AB}$ the mutual information between Alice and Bob,
and $I_{AE}$ the mutual information between Alice and Eve. In what follows we will prepare the ground for defining
and computing those mutual informations for our problem. Also, since the present teleportation-based CVQKD protocol is symmetric to both 
matching conditions, we will work with the one where Alice teleported a real state and Bob implemented the real displacement.

Let $X$ and $Y$ be two binary discrete variables, whose possible values for $X$ are $x=0,1$ and for $Y$ are $y=0,1$. If we associate variable 
$X$ to Alice and adopt the convention $\{-|\alpha\rangle,|\alpha\rangle\}=\{0,1\}$ we have
\begin{equation}
P_X(0) = P_X(1) = 1/2,
\label{pa}
\end{equation}
where $P_X(x)$ is the probability distribution associated to $X$. 
This means that Alice randomly chooses between the negative or positive coherent states at each run of the protocol.

If we associate variable $Y$ to Bob we can define the conditional probability of Bob assigning the value $y$ to his
variable if Alice assigned the value $x$ as $P_{Y|X}(y|x)$. For the present protocol, and according to the encoding that Alice and
Bob mutually agreed on for the key, the four conditional probabilities are 
\begin{eqnarray}
P_{Y|X}(0|0) &=& q_0^B(-\alpha), \\
P_{Y|X}(1|0) &=& 1 - q_0^B(-\alpha), \\
P_{Y|X}(0|1) &=& q_0^B(\alpha), \\
P_{Y|X}(1|1) &=& 1-q_0^B(\alpha),
\end{eqnarray}
where $q_0^B$, the probability to detect the vacuum state, is given by equation (\ref{q0}). If we define 
\begin{equation}
q_1^B(\alpha)=1 - q_0^B(\alpha),
\end{equation}
where $q_1^B$ is the probability to detect light, we have 
\begin{eqnarray}
P_{Y|X}(0|0) &=& q_0^B(-\alpha), \\
P_{Y|X}(1|0) &=& 1 - q_0^B(-\alpha), \\
P_{Y|X}(0|1) &=& 1-q_1^B(\alpha), \\
P_{Y|X}(1|1) &=& q_1^B(\alpha).
\end{eqnarray}
Note that we have explicitly written the dependence of $q_j^B$, $j=0,1$, on Alice's teleported state to remind us that 
we should compute it  using the appropriate sign for $\alpha$. 

We can understand the previous conditional probabilities as
follows. If Alice teleports the state $|-\alpha\rangle$ (bit $0$) and Bob displaces his mode by $\alpha$, 
for a faithful teleportation he will likely detect the vacuum state after that final displacement and assign correctly the bit $0$. 
The chance for that happening is quantified by $P_{Y|X}(0|0) = q_0^B(-\alpha)$. He will obviously make a mistake, assigning erroneously the bit $1$, if he 
does not detect the vacuum state. For that reason we have $P_{Y|X}(1|0) = 1 - q_0^B(-\alpha)$.
In the same fashion, if Alice teleports the state $|\alpha\rangle$ (bit $1$) and Bob displaces his mode by $\alpha$, 
for a faithful teleportation he will very likely not detect the vacuum state and will correctly assign the bit $1$. 
This event occurs with probability $1-q_0^B(\alpha)$, which implies $P_{Y|X}(1|1) = 1-q_0^B(\alpha)$.
He makes a mistake if he gets the vacuum state and therefore $P_{Y|X}(0|1) = q_0^B(\alpha)$.

Since the conditional probability is related to the joint probability distribution $P_{XY}(x,y)$ by the rule
$P_{XY}(x,y)=P_X(x)P_{Y|X}(y|x)$ we have
\begin{eqnarray}
P_{XY}(0,0) &=& q_0^B(-\alpha)/2, \label{pab1}\\
P_{XY}(0,1) &=& [1 - q_0^B(-\alpha)]/2, \\
P_{XY}(1,0) &=& [1-q_1^B(\alpha)]/2, \\
P_{XY}(1,1) &=& q_1^B(\alpha)/2.\label{pab4}
\end{eqnarray}
If we now use that $P_Y(y)=\sum_{x}P_{XY}(x,y)$ we have
\begin{eqnarray}
P_{Y}(0) &=& [1+q_0^B(-\alpha)-q_1^B(\alpha)]/2, \\
P_{Y}(1) &=& [1+q_1^B(\alpha)-q_0^B(-\alpha)]/2. \label{pb2}
\end{eqnarray}

The mutual information between Alice and Bob is defined as
\begin{equation}
I_{AB} = \sum_{x=0}^1\sum_{y=0}^1P_{XY}(x,y)\log_2\left[\frac{P_{XY}(x,y)}{P_X(x)P_Y(y)}\right]
\end{equation}
and a direct computation using Eqs.~(\ref{pa}) and (\ref{pab1})-(\ref{pb2}) gives
\begin{eqnarray}
I_{AB} &=& 1 +[q_0^B\log_2q_0^B+(1-q_0^B)\log_2(1-q_0^B)
+q_1^B\log_2q_1^B\nonumber \\
&&\hspace{-0.5cm}+(1-q_1^B)\log_2(1-q_1^B)]/2 -[(1+q_0^B-q_1^B)\log_2(1+q_0^B-q_1^B) \nonumber \\
&&+ (1+q_1^B-q_0^B)\log_2(1+q_1^B-q_0^B)]/2. 
\label{iab}
\end{eqnarray}
Here we have dropped the $\pm \alpha$ dependence since $q_0^B$ is always computed with $-\alpha$ and
$q_1^B$ with $\alpha$. Note that $I_{AB}$ also  depends on $r,\theta, g_v, g_u$, and $\eta$.
In order to obtain $I_{AE}$ we simply replace $\eta$ for $1-\eta$ in the expression 
for $I_{AB}$ since $q_j^B \rightarrow q_j^E$ if $\eta \rightarrow 1-\eta$. 

Using equation (\ref{iab}) and the equivalent
one for $I_{AE}$ we can compute the secret key rate $K$ (equation (\ref{k})). 
Figure 4 in the main
text was obtained this way, where we employed for each curve a different value for $\eta$ and for all of them 
the optimal values of $r,\theta, g_v$, and $g_u$ assuming the real matching condition as given in figure 2 of the main text.

Note that when the loss is precisely $50\%$ no key can be extracted since Bob's and Eve's state are exactly the same, leading 
to $I_{AB}=I_{AE}$ and $K=0$.
When the loss is exactly $100\%$, the protocol does not work either. In this case Bob's state is the vacuum state $|0\rangle$, i.e.,
he receives no signal, and Eve can also operate on a vacuum state instead of the intercepted signal. It is clear, thus, that
Bob and Eve will have the same mutual information with Alice and obviously $K=0$. 

This suggests a possible attack on the
present protocol whenever we have high losses. Indeed, Eve can work with a vacuum state instead of her share of the 
intercepted signal since the former is closer to the state with Bob, whose state in a very lossy
environment is nearly the vacuum state. Therefore, we have to improve the security analysis when we have great losses in order to handle
the fact that Eve can work with both the intercepted signal and the vacuum state. In this situation, 
the effective secure key rate that can be achieved between Alice and Bob is
\begin{equation}
K_e = \max\{0,\min\{\Delta I,\Delta I_0\}\},
\label{k2}
\end{equation}
where $\Delta I_0 = \beta I_{AB}-I^0_{AE}$ and $I^0_{AE}$ is the mutual information between Alice and Eve 
assuming Eve's state is the vacuum. $I^0_{AE}$ is easily obtained from the general expression for $I_{AE}$
by setting $\eta=1.0$, the case where Bob receives the whole signal and Eve gets nothing, i.e., she has
the vacuum state.

Table I of the main text was obtained maximizing $K_e$ for several values of fixed $r$, $\eta$, and $\beta$. 
Equations (\ref{gur}) and (\ref{gvr}) was used for $g_u$ and $g_v$ and $\theta$ was determined in such a way that $K_e$ be maximal. 
As always, we assumed the real matching condition to fix the remaining parameters needed to evaluate $K_e$, namely, Alice's input was either $|-\alpha \rangle$ 
or $|\alpha \rangle$ and Bob's final displacement was $\hat{D}(\alpha)$.

\section{Further examples}
\label{E}

\begin{figure}[!ht]
\begin{center}
\includegraphics[width=8cm]{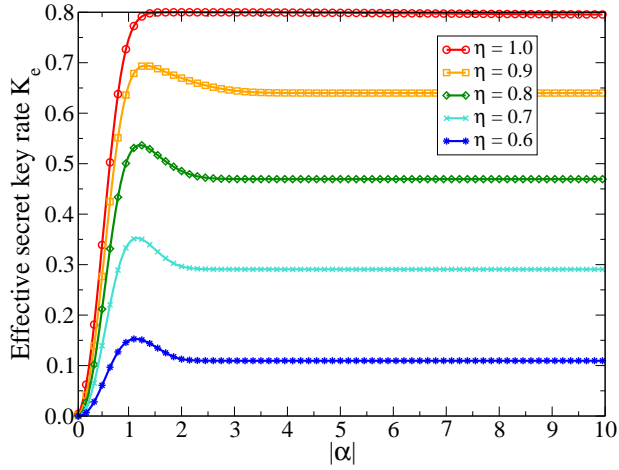}
\end{center}
\caption{ \label{FigOptA} Here both $r$ and $\theta$ are adjusted to get the optimal key rates with $\beta=0.8$.
We assume the real matching condition.
}
\end{figure}
\begin{figure}[!ht]
\begin{center}
\includegraphics[width=8cm]{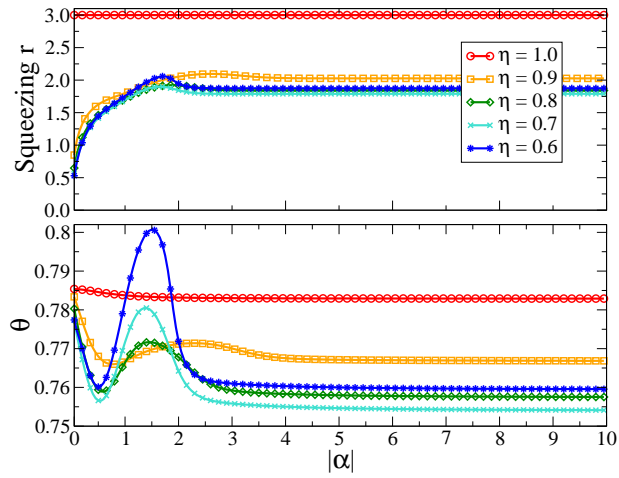}
\end{center}
\caption{ \label{FigParA} 
Optimal parameters leading to the key rates in figure \ref{FigOptA}. In the maximization process we have restricted $r$ from $0$ to $3$ while
$\theta$ could assume any value. The optimal $g_u$ and $g_v$ are obtained using these values of $\theta$ and $r$ to evaluate
Eqs. (\ref{gur}) and (\ref{gvr}). 
}
\end{figure}

\begin{figure}[!ht]
\begin{center}
\includegraphics[width=8cm]{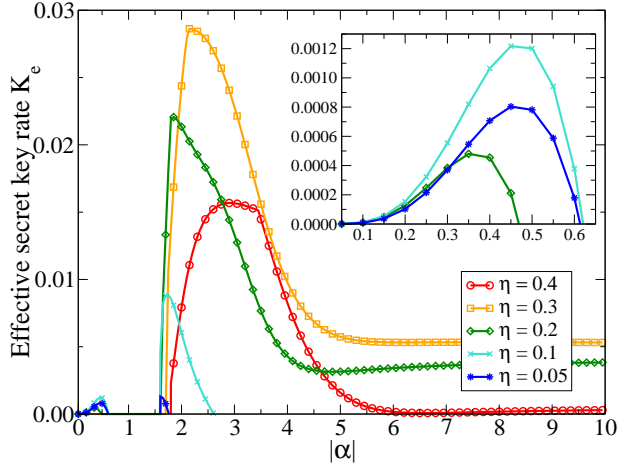}
\end{center}
\caption{ \label{FigOptB} Here both $r$ and $\theta$ are adjusted to get the optimal key rates with $\beta=0.8$.
We assume the real matching condition.
}
\end{figure}
\begin{figure}[!ht]
\begin{center}
\includegraphics[width=8cm]{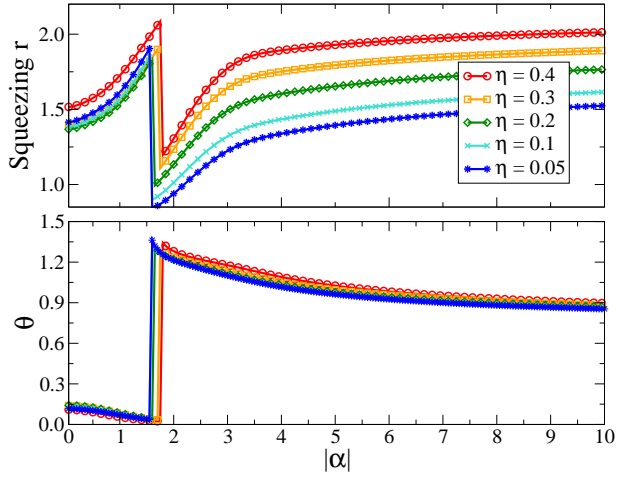}
\end{center}
\caption{ \label{FigParB} 
Optimal parameters leading to the key rates in figure \ref{FigOptB}. In the maximization process we have restricted $r$ from $0$ to $3$ while
$\theta$ could assume any value. The optimal $g_u$ and $g_v$ are obtained using these values of $\theta$ and $r$ to evaluate
Eqs. (\ref{gur}) and (\ref{gvr}). 
}
\end{figure}

Assuming squeezing is a cheap resource, we can let $r$, together with $\theta$, be a free parameter in the maximization of the key rate.
In this scenario, we get the results in Figs. \ref{FigOptA} and \ref{FigOptB} for the effective optimal key rates for several values of loss. 
The optimal parameters leading to such key rates are given in Figs. \ref{FigParA} and \ref{FigParB}. 



It is interesting to note that whenever we
have loss ($\eta \neq 1.0$) the optimal squeezing is not the greatest value possible.  
For losses lower than $50\%$ ($\eta$ from $1.0$ to $0.6$) the greater the loss the lower the key rate. Interestingly, the behavior for 
losses greater than $50\%$ is different. Once you cross the border of $50\%$ loss, more loss means a better key rate. But this trend stops at
about $70\%$ loss ($\eta = 0.3$), from which the key rate starts to decrease again with loss. When the exact values of
$50\%$ or $100\%$ loss is used, no effective key rate can be achieved since Bob and Eve share the same level of 
information with Alice. We also remark that in most of the cases the optimal squeezing is not greater than $r=2.0 (17.4 \mbox{dB})$.

Finally, it is important to note that for losses lower than $50\%$, i.e., when more than half of the signal sent from Alice reaches Bob, the
effective key rate $K_e$ is simply $K$ as given by equation (\ref{k}). However, when we go beyond the level of $50\%$ loss, 
equation (\ref{k2}) starts to be relevant. Depending on the value of $|\alpha|$, either $\Delta I$ or $\Delta I_0$ is the lowest term
that defines the key. That is why in the cases with more than $50\%$ loss the curves for $K_e$ have an abrupt behavior.
And for very high loss, $\Delta I_0$ is always the lowest term.


\begin{thebibliography}{200}

\bibitem{Ben84}  Bennett, C. H., Brassard, G.: Quantum cryptography: public key distribution and coin tossing. In: IEEE
International Conference on Computers, Systems and Signal Processing, pp. 175 (1984)

Ekert, A. K.: Quantum cryptography based on Bell’s theorem. Phys. Rev. Lett. \textbf{67}, 661 (1991);
Bennett, C. H.: Quantum cryptography using any two nonorthogonal states. Phys. Rev. Lett. \textbf{68}, 3121 (1992); 
Bennett, C. H., Brassard, G., Mermin, N. D.: Quantum cryptography without Bell’s theorem. Phys. Rev. Lett. \textbf{68}, 557 (1992)

\bibitem{Gis02} Extensive reviews of the discrete variable protocols in \cite{Ben84} and their descendants can be found in: 
Gisin, N., Ribordy, G., Tittle, W., Zbinden, H.: Quantum cryptography. Rev. Mod. Phys. \textbf{74}, 145 (2002); 
Scarani, V., Bechmann-Pasquinucci, H., Cerf, N. J., Du\v{s}ek, M., L\"utkenhaus, N., Peev, M.: The security of practical quantum key distribution.
Rev. Mod. Phys. \textbf{81}, 1301 (2009)

\bibitem{Ral99} Ralph, T. C.: Continuous variable quantum cryptography. Phys. Rev. A \textbf{61}, 010303(R) (1999); 
Hillery, M.:  Quantum cryptography with squeezed states. Phys. Rev. A \textbf{61}, 022309 (2000);
Reid, M. D.: Quantum cryptography with a predetermined key, using continuous-variable Einstein-Podolsky-Rosen correlations. Phys. Rev. A \textbf{62}, 062308 (2000)

\bibitem{Cer01} Cerf, N. J., L\'evy, M., Van Assche, G.: Quantum distribution of Gaussian keys using squeezed states. Phys. Rev. A \textbf{63}, 052311 (2001)

\bibitem{Gro02} Grosshans, F., Grangier, Ph.: Continuous Variable Quantum Cryptography Using Coherent States. Phys. Rev. Lett. \textbf{88}, 057902 (2002)

\bibitem{Leu02} Silberhorn, Ch.,  Ralph, T. C., L\"utkenhaus, N., Leuchs, G.: Continuous Variable Quantum Cryptography: Beating the 3 dB Loss Limit. Phys. Rev. Lett. \textbf{89}, 167901 (2002);
Lorenz, S., Korolkova, N., Leuchs, G.: Continuous-variable quantum key distribution using polarization encoding and post selection. Appl. Phys. B \textbf{79}, 273 (2004)

\bibitem{Gro03} Grosshans, F., Van Assche, G., Wenger, J., Brouri, R., Cerf, N. J., Grangier, Ph.: Quantum key distribution using gaussian-modulated coherent states. Nature (London) \textbf{421}, 238 (2003);
Legr\'e, M., Zbinden, H., Gisin, N.: Implementation of continuous variable quantum cryptography in optical fibres using a go-\&-return configuration. Quantum Inf. Comput. \textbf{6}, 326 (2006);
Lodewyck, J., Debuisschert, T., Tualle-Brouri, R., Grangier, Ph.: Controlling excess noise in fiber-optics continuous-variable quantum key distribution. Phys. Rev. A \textbf{72}, 050303(R) (2005);
Lodewyck, J., Bloch, M., Garc\'ia-Patr\'on, R., Fossier, S., Karpov, E., Diamanti, E., Debuisschert, T.,
Cerf, N. J., Tualle-Brouri, R., McLaughlin, S. W., Grangier, Ph.: Quantum key distribution over 25km with an all-fiber continuous-variable system. Phys. Rev. A \textbf{76}, 042305 (2007);
Jouguet, P., Kunz-Jacques, S., Leverrier, A., Grangier, Ph., Diamanti, E.: Experimental demonstration of long-distance continuous-variable quantum key distribution. Nature Photonics \textbf{7}, 378 (2013)

\bibitem{Nam03} Hirano. T., Yamanaka, H., Ashikaga, M., Konishi, T., Namiki, R.: Quantum cryptography using pulsed homodyne detection. Phys. Rev. A \textbf{68}, 042331 (2003); 
Namiki, R., Hirano, T.: Security of quantum cryptography using balanced homodyne detection. Phys. Rev. A \textbf{67}, 022308 (2003); 
Namiki, R., Hirano, T.: Practical Limitation for Continuous-Variable Quantum Cryptography using Coherent States. Phys. Rev. Lett. \textbf{92}, 117901 (2004);
Namiki, R., Hirano, T.: Efficient-phase-encoding protocols for continuous-variable quantum key distribution using coherent states and postselection. Phys. Rev. A \textbf{74}, 032302 (2006)

\bibitem{Wee04} Weedbrook, Ch., Lance,A. M., Bowen, W. P., Symul, Th., Ralph, T. C., Lam, P. K.: Quantum Cryptography Without Switching. Phys. Rev. Lett. \textbf{93}, 170504 (2004);
Lance, A. M., Symul, Th., Sharma, V., Weedbrook, Ch., Ralph, T. C., Lam, P. K.: No-Switching Quantum Key Distribution Using Broadband Modulated Coherent Light. Phys. Rev. Lett. \textbf{95}, 180503 (2005)

\bibitem{Hei06} Heid, M., L\"utkenhaus, N.: Efficiency of coherent-state quantum cryptography in the presence of loss: Influence of realistic error correction. Phs. Rev. A \textbf{73} 052316 (2006);
Heid, M., L\"utkenhaus, N.: Security of coherent-state quantum cryptography in the presence of Gaussian noise. Phs. Rev. A \textbf{76} 022313 (2007)

\bibitem{Pir08} Pirandola, S., Mancini, S., Lloyd, S., Braunstein, S. L.: Continuous-variable quantum cryptography using two-way quantum communication. Nature Phys. \textbf{4}, 726 (2008)

\bibitem{Pat09} Garc\'ia-Patr\'on, R., Cerf, N. J.: Continuous-Variable Quantum Key Distribution Protocols Over Noisy Channels. Phys. Rev. Lett. \textbf{102}, 130501 (2009)

\bibitem{Lev09} Leverrier, A., Grangier, Ph.: Unconditional Security Proof of Long-Distance Continuous-Variable Quantum Key Distribution with Discrete Modulation. Phys. Rev. Lett. \textbf{102}, 180504 (2009); 
Leverrier, A., Grangier, Ph.: Continuous-variable quantum-key-distribution protocols with a non-Gaussian modulation. Phys. Rev. A \textbf{83}, 042312 (2011)

\bibitem{Leu10} Sych, D., Leuchs, G.: Coherent state quantum key distribution with multi letter phase-shift keying. New J. Phys. \textbf{12}, 053019 (2010)

\bibitem{Mad12} Madsen, L. S., Usenko, V. C., Lassen, M., Filip, R., Andersen, U. L.: Continuous variable quantum key distribution with modulated entangled states. Nat. Commun. 3:1083 doi: 10.1038/ncomms2097 (2012)

\bibitem{Pir14} Pirandola, S., Ottaviani, C., Spedalieri, G., Weedbrook, Ch., Braunstein, S. L., Lloyd, S.,
Gehring, T., Jacobsen, Ch. S., Andersen, U. L.: High-rate measurement-device-independent quantum cryptography. Nature Photonics \textbf{9}, 397 (2015); 
Li, Z., Zhang, Y.-C., Xu, F., Peng, X., Guo,H.: Continuous-variable measurement-device-independent quantum key distribution. Phys. Rew. A \textbf{89}, 052301 (2014)

\bibitem{Bor16} Borelli, L. F. M., Aguiar, L. S., Roversi, J. A., Vidiella-Barranco, A.: Quantum key distribution using continuous-variable non-Gaussian states. Quantum Inf. Process. \textbf{15}, 893 (2016)

\bibitem{Bra05} See \cite{Gis02} and in particular the following references for reviews on CVQKD protocols: 
Braunstein, S. L., van Loock, P.: Quantum information with continuous variables. Rev. Mod. Phys. \textbf{77}, 513 (2005);
Weedbrook, Ch., Pirandola, S., Garc\'ia-Patr\'on, R., Cerf, N. J., Ralph, T. C., Shapiro, J. H., Lloyd, S.: Gaussian quantum information. Rev. Mod. Phys. \textbf{84}, 621 (2012)

\bibitem{Gro04} Gottesman, D., Preskill, J.: Secure quantum key distribution using squeezed states. Phys. Rev. A \textbf{63}, 022309 (2001);
Grosshans, F., Cerf, N. J.: Continuous-Variable Quantum Cryptography is Secure against Non-Gaussian Attacks. Phys. Rev. Lett. \textbf{92}, 047905 (2004); 
Iblisdir, S., Van Assche, G., Cerf, N. J.: Security of Quantum Key Distribution with Coherent States and Homodyne Detection. Phys. Rev. Lett. \textbf{93}, 170502 (2004);
Grosshans, F.: CollectiveAttacks and Unconditional Security in Continuous Variable Quantum KeyDistribution. Phys. Rev. Lett. \textbf{94}, 020504 (2005);  
Navascu\'es, M., Ac\'in, A.: SecurityBounds for Continuous Variables Quantum Key Distribution. Phys. Rev. Lett. \textbf{94}, 020505 (2005); 
Navascu\'es, M., Grosshans, F., Ac\'in, A.: Optimality of Gaussian Attacks in Continuous-Variable Quantum Cryptography. Phys. Rev. Lett. \textbf{97}, 190502 (2006); 
Garc\'ia-Patr\'on, R., Cerf, N. J.: Unconditional Optimality of Gaussian Attacks against Continuous-Variable Quantum Key Distribution. Phys. Rev. Lett. \textbf{97}, 190503 (2006);
Renner, R., Cirac, J. I.: de Finetti Representation Theorem for Infinite-Dimensional Quantum Systems and Applications to Quantum Cryptography. Phys. Rev. Lett. \textbf{102}, 110504 (2009);
Zhao, Y.-B., Heid, M., Rigas, J., L\"utkenhaus, N.: Asymptotic security of binary modulated continuous-variable quantum key distribution under collective attacks. Phys. Rev. A \textbf{79}, 012307 (2009);
Weedbrook, Ch., Pirandola, S., Lloyd, S., Ralph, T. C.: Quantum Cryptography Approaching the Classical Limit. Phys. Rev. Lett. \textbf{105}, 110501 (2010);
Leverrier, A., Garc\'ia-Patr\'on, R., Renner, R., Cerf, N. J.: Security of Continuous-Variable Quantum Key Distribution Against General Attacks. Phys. Rev. Lett. \textbf{110}, 030502 (2013);
Jouguet, P., Kunz-Jacques, S., Diamanti, E.: Preventing calibration attacks on the local oscillator in continuous-variable quantum key distribution. Phys. Rev. A \textbf{87}, 062313 (2013);
Huang, J.-Z., Kunz-Jacques, S., Jouguet, P., Weedbrook, Ch., Yin, Z.-Q., Wang, Sh., Chen, W., Guo, G.-C., Han, Z.-F.: Quantum hacking on quantum key distribution using homodyne detection. Phys. Rev. A \textbf{89}, 032304 (2014)

\bibitem{Bra98} Vaidman, L.: Teleportation of quantum states. Phys. Rev. A \textbf{49}, 1473 (1994); 
Braunstein, S. L., Kimble, H. J.: Teleportation of Continuous Quantum Variables. Phys. Rev. Lett. \textbf{80}, 869 (1998);
Furusawa, A., S{\o}rensen, J. L., Braunstein, S. L., Fuchs, C. A., Kimble, H. J., Polzik, E. S.: Unconditional Quantum Teleportation. Science \textbf{282}, 706 (1998)

\bibitem{Fur07} Yoshino, K.-i., Aoki, T., Furusawa, A.: Generation of continuous-wave broadband entangled beams using periodically poled lithium niobate waveguides. Appl. Phys. Lett. \textbf{90}, 041111 (2007);
Lee, N., Benichi, H., Takeno, Y., Takeda, Sh., Webb, J., Huntington, E., Furusawa, A.: Teleportation of Nonclassical Wave Packets of Light. Science \textbf{332}, 330 (2011)

\bibitem{Gor10} Gordon, G., Rigolin, G.: Quantum cryptography using partially entangled states. Opt. Commun. \textbf{283}, 184 (2010)

\bibitem{Lui14} Luiz, F. S., Rigolin, G.: Optimal continuous variable quantum teleportation protocol for realistic settings. Annals of Physics \textbf{354}, 409 (2015)

\bibitem{Bec13} Becir, A., Wahiddin, M. R. B.: Tight bounds for the eavesdropping collective attacks on general CV-QKD protocols that involve non-maximally entanglement.
Quantum Inf. Process. \textbf{12}, 1155 (2013)

\end{thebibliography}
\end{document}